\documentclass{aa}  

\usepackage{graphicx}
\usepackage{txfonts}
\usepackage[colorlinks=true,citecolor=blue,linkcolor=blue,urlcolor=blue,breaklinks=true]{hyperref}
\usepackage{xcolor,float}
\usepackage[ruled,vlined]{algorithm2e}
\usepackage{multirow}
\usepackage{wasysym}
\usepackage{gensymb}
\usepackage{natbib}
\bibpunct{(}{)}{;}{a}{}{,}

\newcommand{\FRB}{FRB\,20201020A}
\newcommand{\RIII}{FRB\,20180916B}
\newcommand{\RI}{FRB\,20121102A}
\newcommand{\FRBperiodic}{FRB\,20191221A}
\newcommand{\FRBfivecomp}{FRB\,20210206A}
\newcommand{\FRBsixcomp}{FRB\,20210213A}
\newcommand{\chimefrbs}{CHIME FRBs 20210206A and 20210213A}
\newcommand{\jyms}{\,Jy\,ms} % Fluence units
\newcommand{\radsqm}{\,rad\,m$^{-2}$} % Rotation measure units
\newcommand{\pccm}{\,pc\,cm$^{-3}$} % Dispersion measure units
\newcommand{\us}{\,$\mu$s} % Microseconds
\newcommand{\ra}[3]{$#1^{\text{h}}#2^{\text{m}}#3^{\text{s}}$}
\newcommand{\dec}[3]{$#1\degr#2\arcmin#3\arcsec$}
\newcommand{\rchisq}{r$\chi^2$}
\newcommand{\schisq}{$\sigma_{\chi^2}$}
\newcommand{\sshat}{$\sigma_{\hat{S}}$}
\newcommand{\Psc}{$P_{\text{sc}}$}

\iftrue  % Use this to show comments
    \def\ipm#1{\textcolor{teal}{\bf[#1 -- IPM]}\xspace}
    \def\jvl#1{\textcolor{orange}{\bf[#1 -- JVL]}\xspace}
    \def\ab#1{\textcolor{purple}{\bf[#1 -- AB]}\xspace}
    \def\ym#1{\textcolor{brown}{\bf[#1 -- YM]}\xspace}
    \def\prev#1{\textcolor{blue}{\bf[was: #1 -- IPM]}\xspace}
    \def\sms#1{\textcolor{magenta}{\bf[#1 -- SMS]}\xspace}   
\else
  \def\ipm#1{}
  \def\jvl#1{}
  \def\ab#1{}
  \def\ym#1{}
  \def\sms#1{}
  \def\prev#1{}
  
\fi

\begin{document}

   \title{A fast radio burst with sub-millisecond quasi-periodic structure}

\author{%
     Inés~Pastor-Marazuela \inst{\ref{uva} \and \ref{astron}}
\and Joeri~van~Leeuwen    \inst{\ref{astron} \and \ref{uva}}
\and Anna~Bilous          \inst{\ref{astron}}
\and Liam~Connor          \inst{\ref{caltech} \and \ref{uva}}
\and Yogesh~Maan          \inst{\ref{ncra} \and \ref{astron}}
\and Leon~Oostrum         \inst{\ref{astron} \and \ref{uva} \and \ref{escience}}
\and Emily~Petroff        \inst{\ref{uva} \and \ref{veni} \and \ref{mu}}
\and Samayra~Straal       \inst{\ref{nyuad} \and \ref{nyuadcfa}}
\and Dany~Vohl            \inst{\ref{uva} \and \ref{astron}}
\and E.~A.~K.~Adams       \inst{\ref{astron} \and \ref{kapteyn}}
\and B.~Adebahr           \inst{\ref{airub}}
\and Jisk~Attema          \inst{\ref{escience}}
\and Oliver~M.~Boersma \inst{\ref{uva} \and \ref{astron}}
\and R.~van~den~Brink     \inst{\ref{astron}}
\and W.A.~van~Cappellen   \inst{\ref{astron}}
\and A.~H.~W.~M.~Coolen   \inst{\ref{astron}}
\and S.~Damstra           \inst{\ref{astron}}
\and H.~Dénes            \inst{\ref{astron}}
\and K.~M.~Hess           \inst{\ref{iaa} \and \ref{astron} \and \ref{kapteyn}}
\and J.~M.~van~der~Hulst  \inst{\ref{kapteyn}}
\and B.~Hut               \inst{\ref{astron}}
\and A.~Kutkin            \inst{\ref{astron}}
\and G.~Marcel~Loose      \inst{\ref{astron}}
\and D.~M.~Lucero         \inst{\ref{virginiatech}}
\and {\'A}.~Mika             \inst{\ref{astron}}
\and V.~A.~Moss           \inst{\ref{csiro} \and \ref{sydney} \and \ref{astron}}
\and H.~Mulder            \inst{\ref{astron}}
\and M.J.~Norden          \inst{\ref{astron}}
\and T.~A.~Oosterloo      \inst{\ref{astron} \and \ref{kapteyn}}
\and Kaustubh~Rajwade     \inst{\ref{astron}}
\and D.~van~der~Schuur    \inst{\ref{astron}}
\and A.~Sclocco           \inst{\ref{escience}}
\and R.~Smits             \inst{\ref{astron}}
\and J.~Ziemke            \inst{\ref{astron} \and \ref{oslocit}}
}

%% ----------------- Institutes --------------- 
\institute{Anton Pannekoek Institute, University of Amsterdam, Postbus 94249, 1090 GE Amsterdam, The Netherlands\label{uva}
  \and
ASTRON, the Netherlands Institute for Radio Astronomy, Oude Hoogeveensedijk 4,7991 PD Dwingeloo, The Netherlands\label{astron}
  \and
Cahill Center for Astronomy, California Institute of Technology, Pasadena, CA, USA\label{caltech}
  \and
National Centre for Radio Astrophysics, Tata Institute of Fundamental Research, Pune 411007, Maharashtra, India\label{ncra}
  \and
Netherlands eScience Center, Science Park 140, 1098 XG, Amsterdam, The Netherlands\label{escience}
  \and
Veni Fellow\label{veni}
  \and
Department of Physics, McGill University, 3600 rue University, Montr{\'e}al, QC H3A 2T8, Canada\label{mu}
  \and
NYU Abu Dhabi, PO Box 129188, Abu Dhabi, United Arab Emirates\label{nyuad}
  \and
Center for Astro, Particle, and Planetary Physics (CAP$^3$), NYU Abu Dhabi, PO Box 129188, Abu Dhabi, United Arab Emirates\label{nyuadcfa}
  \and
Kapteyn Astronomical Institute, PO Box 800, 9700 AV Groningen, The Netherlands\label{kapteyn}
  \and
Astronomisches Institut der Ruhr-Universit{\"a}t Bochum (AIRUB), Universit{\"a}tsstrasse 150, 44780 Bochum, Germany\label{airub}
  \and
Instituto de Astrofísica de Andalucía (CSIC), Glorieta de la Astronomía s/n, 18008 Granada, Spain\label{iaa}
  \and
Department of Physics, Virginia Polytechnic Institute and State University, 50 West Campus Drive, Blacksburg, VA 24061, USA\label{virginiatech}
  \and
CSIRO Astronomy and Space Science, Australia Telescope National Facility, PO Box 76, Epping NSW 1710, Australia\label{csiro}
  \and
Sydney Institute for Astronomy, School of Physics, University of Sydney, Sydney, New South Wales 2006, Australia\label{sydney}
  \and
University of Oslo Center for Information Technology, P.O. Box 1059, 0316 Oslo, Norway\label{oslocit}}

%   \author{I. Pastor-Marazuela\inst{1,2}
%          \and
%          J. van Leeuwen\inst{2,1}
%          \and
%%          ARTS Team
 %         \and
 %         Apertif builders
 %         }
%   \institute{Anton Pannekoek Institute for Astronomy, University of Amsterdam, Science Park 904, 1098 XH Amsterdam, The Netherlands\\
%              \email{i.pastormarazuela@uva.nl }
%         \and
%             ASTRON, Netherlands Institute for Radio Astronomy, Oude Hoogeveensedijk 4, 7991 PD Dwingeloo, The Netherlands
%             }

   \date{}

% \abstract{}{}{}{}{} 
% 5 {} token are mandatory
 
  \abstract
  % context heading (optional)
  {Fast radio bursts (FRBs) are extragalactic radio transients of extraordinary luminosity. 
  Studying the diverse temporal and spectral behaviour recently observed in a number of FRBs may help determine the nature of the entire class. 
  For example, a fast spinning or highly magnetised neutron star might generate the rotation-powered acceleration required to explain the bright emission.
  Periodic, sub-second components, suggesting such rotation, were recently reported in one FRB, and potentially in two more.
   }
  % aims heading (mandatory)
   {Here we report the discovery of \FRB{} with  Apertif, an FRB that shows five components regularly spaced by 0.415\,ms. 
   This sub-millisecond structure in \FRB{} carries important clues about the progenitor of this FRB specifically, and potentially about that of FRBs in general.
   We thus contrast its features to the predictions of the main FRB source models.
      }
  % methods heading (mandatory)
   {We perform a timing analysis of the  \FRB{} components to determine the significance of the periodicity. We compare these against the timing properties of the previously reported CHIME FRBs with sub-second quasi-periodic components, and against  two Apertif bursts from  
   repeating \RIII{} that show complex time-frequency structure.
   }
  % results heading (mandatory)
   {We find the periodicity of \FRB{} to be marginally significant at 2.5$\sigma$. Its repeating subcomponents cannot be explained as a pulsar rotation since the required spin rate of over 2\,kHz exceeds the limits set by typical neutron star equations of state and observations. The fast periodicity is also in conflict with a compact object merger scenario. These quasi-periodic components could, however, be caused by equidistant emitting regions in the magnetosphere of a magnetar. 
  }
  % conclusions heading (optional), leave it empty if necessary
   {
   The sub-millisecond spacing of the components in \FRB{}, the smallest observed so far in a one-off FRB, may rule out both neutron-star rotation and binary mergers as the direct source of quasi-periodic FRBs.
}


   \maketitle
%
%-------------------------------------------------------------

\section{Introduction}
\label{sec:introduction}

Fast radio bursts (FRBs) are bright, millisecond-duration radio transients of extragalactic origin that have puzzled
researchers since their discovery 
(\citealt{lorimer_bright_2007}; see \citealt{petroff_fast_2019} and \citealt{cordes_fast_2019} for a review of properties).
While most FRBs are only seen once (one-offs), around 24 %of them
%\jvl{``of them'' can almost always be deleted. I propose you do a search-replace. Here too, "around 24 have been seen'' is better } 
have been seen to repeat \citep[e.g.][]{spitler_repeating_2016,the_chime/frb_collaboration_second_2019}. There is no consensus yet on whether these two observational classes are produced by the same types of sources or if they have different origins. %\sms{they could still have the same progenitor, but different origin.} 
Many models invoke compact objects as the source of FRBs \citep[see][]{platts_living_2019}, and several observational clues point at magnetars as the progenitors of at least some FRBs \citep[see][and references therein]{zhang_physical_2020}.
After the detection of a bright radio burst from the Galactic magnetar SGR\,1935+2154 \citep{bochenek_fast_2020, the_chimefrb_collaboration_bright_2020}, we now know that at least some FRBs could be produced by magnetars.

Recently, the Canadian Hydrogen Intensity Mapping Experiment Fast Radio Burst Project (CHIME/FRB) published the largest catalog of FRBs detected with a single instrument to date \citep{the_chimefrb_collaboration_first_2021}. While repeaters and non repeaters show similar sky distributions, dispersion measures (DM) and scattering timescales, repeaters have been seen to display a distinctive time and frequency structure commonly referred to as the "sad trombone effect" \citep{hessels_frb_2019}, in which multi-component bursts drift downwards in frequency. On the other hand, one-off FRBs can present single-component bursts, either narrow or broadband, as well as multi-component bursts with similar frequency extent \citep{pleunis_fast_2021}.

Several works have carried out short-timescale timing analyses, including periodicity searches, in multi-component bursts from repeating FRBs. \cite{nimmo_highly_2021, nimmo_burst_2021} find sub-component separations in \RIII{} and FRB\,20200120E of just a few $\mu$s, but no evidence of periodicity. Meanwhile, \cite{majid_bright_2021} find hints for a regular separation of 2-3\,$\mu$s in the subcomponents of a burst from FRB\,20200120E.

\cite{the_chimefrb_collaboration_sub-second_2021} recently reported the detection of a one-off FRB with $\geq$9 components, \FRBperiodic, showing a strict periodic separation of 216.8\,ms between its components and a total duration of \mbox{$\sim3$\,s}. This phenomenon is different from the long-term periodicity that two repeating FRBs have earlier been proved to show; \RIII{} with a period of $\sim16.3$\,days \citep{the_chimefrb_collaboration_periodic_2020}, and \RI{} with a period of $\sim160$\,days \citep{rajwade_possible_2020, cruces_repeating_2020}. These two repeating FRBs show a periodicity in their activity cycles, with a period $>10$\,days, while \FRBperiodic{} shows a periodic structure within the subcomponents of the only detected burst, with a period $<1$\,s. From here on, our use of the term ``periodic'' only refers to this latter, generally sub-second, fast periodicity. %\jvl{In\'es can you check that that is true?}
%However, \FRBperiodic{} represents the first-ever detection of a periodicity within the same burst, and it does not seem to correspond to the same phenomenon as the repeating FRB activity cycles.
\cite{the_chimefrb_collaboration_sub-second_2021} report two additional one-off multi-component FRBs, \FRBfivecomp{}  and
\FRBsixcomp, with apparent periodic separations of 2.8 and 10.7\,ms respectively, though at a lower significance. This
hints to the potential existence of a new sub-group of one-off FRBs showing \mbox{(quasi-)periodic} sub-second components. 

Prompted by the CHIME/FRB detections of \FRBperiodic, \FRBfivecomp{} and \FRBsixcomp, we searched the Apertif FRBs for bursts with (quasi-)periodic structure.
In this work, we report the detection of \FRB{} with the Apertif Radio Transient System \citep[ARTS;][]{leeu14} installed at the Westerbork Synthesis Radio Telescope \citep[WSRT;][]{adams_radio_2019,cov+21}. 
\FRB{} shows five components with a regular spacing of 0.415\,ms.
Furthermore, we perform a detailed timing analysis of the bursts A17 and A53 from \RIII{} that were first reported in \cite{pastor-marazuela_chromatic_2021}, which also exhibit complex time and frequency structure that we compare to that of \FRB.
In Section~\ref{sec:obs} we present the detection, observations and properties of \FRB, including its localisation, scintillation bandwidth and repetition rate upper limit. Section~\ref{sec:results} explains the timing analysis of \FRB{} as well as the \RIII{} bursts A17 and A53. In Section~\ref{sec:discussion} we discuss the interpretation of the temporal structure seen in \FRB, and assert that it belongs to the same morphological type as the \chimefrbs. Finally, we give our conclusions in Section~\ref{sec:conclusions}.

%-------------------------------------------------------------
\section{Observations and data reduction}
\label{sec:obs}
The Apertif Radio Transient System (ARTS) is currently carrying out an FRB survey which started in July 2019 \citep{lko+22} %(van Leeuwen et al. in prep.) 
using ten 25-m dishes of the WSRT. 
In survey observing mode, we beamform 40 compound beams (CBs) covering a field of view (FoV) of 9.5\,deg$^2$. Stokes I data is saved with a time resolution of 81.92\us{} at a central frequency of 1370\,MHz with 300\,MHz bandwidth and a frequency resolution of 
%0.1953125\,MHz 
0.195\,MHz 
\citep[see][for further detail]{maan_real-time_2017,oostrum_repeating_2020}. The Stokes I data is then searched for transients in real-time with \texttt{AMBER}\footnote{\url{https://github.com/TRASAL/AMBER}} \citep{sclocco_real-time_2014,sclocco_real-time_2016,sclocco_real-time_2019}. Next, a machine learning classifier assigns the probability to each candidate of it being a real FRB \citep{connor_applying_2018}, and the best FRB candidates are then sent in an email for human inspection.

\begin{figure}
    \centering
    \includegraphics[width=\hsize]{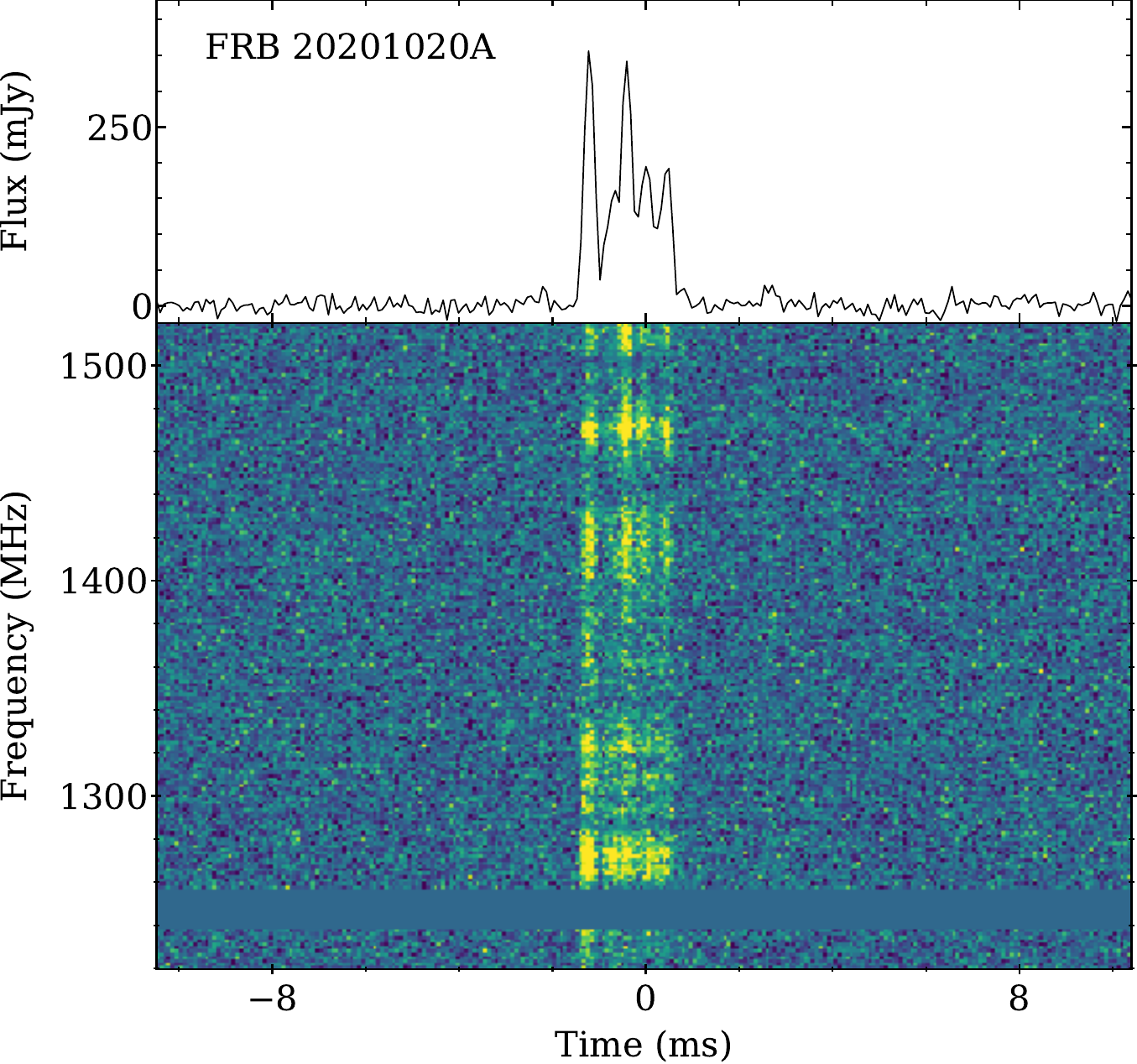}
    \caption{Dynamic spectrum of \FRB, dedispersed to a DM of 398.59\pccm. The top panel shows the average calibrated pulse profile of the burst, and the bottom panel the intensity of the burst in time versus frequency. The data has been downsampled in frequency by a factor 8.}
    \label{fig:frb201020_dynspec}
\end{figure}

\FRB{} was detected during a follow-up observation of the repeating FRB\,20190303A \citep{fonseca_nine_2020}. It was detected in three adjacent compound beams (CBs), with a maximum signal-to-noise ratio (SNR) of $\sim$53. Since its dynamic spectrum shows complex time-frequency structure, we used a structure maximisation algorithm \citep{gajjar_highest-frequency_2018,hessels_frb_2019,pastor-marazuela_chromatic_2021} to get its optimal dispersion measure (DM) of 398.59(8)\pccm. The FRB properties are summarised in Table~\ref{tab:frb201020}, and its dynamic spectrum is shown in Fig.~\ref{fig:frb201020_dynspec}.

The detection of bursts A17 and A53 from \RIII{} was originally reported in \cite{pastor-marazuela_chromatic_2021}, where the observations and data reduction were already explained.

% ARTS has already demonstrated its FRB detection proficiency through the detection of new FRBs with interesting polarisation properties \citep[e.g.][]{connor_bright_2020} as well as the follow-up and characterisation of known repeaters \citep{oostrum_repeating_2020,pastor-marazuela_chromatic_2021}.

% python /home/arts/pastor/scripts/arts-analysis/plot_frbs.py -frb FRB201020 -fluxcal 

\begin{table}
\caption{\FRB{} properties}             % title of Table
\label{tab:frb201020}      % is used to refer this table in the text
\centering                          % used for centering table
\begin{tabular}{l l}        % centered columns (4 columns)
\hline\hline                 % inserts double horizontal lines
\multicolumn{2}{c}{\FRB{} properties} \\    % table heading 
\hline                        % inserts single horizontal line

Dispersion measure (DM) & 398.59(8)\pccm\\
Arrival time & 2020-10-20 12:09:17.385 \\
Barycentric MJD & 59142.50645121 \\
Detection SNR (CB29) & 53.34 \\
Right ascension (J2000) & \ra{13}{51}{25} \\
Declination (J2000) & \dec{+49}{02}{06} \\
Ellipse major and minor axes & $3.9\arcmin\times 17.2\arcsec$ \\
Ellipse angle (N-E) & 103.9$\degr$\\
Fluence & 3.4(7)\jyms\\
Peak flux & 14.0(3)\,Jy \\
Total width & 2.14(2)\,ms \\
Scintillation bandwidth & 9.8(9)\,MHz\\
Rotation measure (RM) & $+110\pm69$\radsqm\\

\hline                                   %inserts single line
\end{tabular}
\end{table}

\begin{table}[]
    \centering
    \caption{Properties of the five \FRB{} components.}
    \begin{tabular}{cll}
    \hline\hline
    Component & ToA (ms) & Width (ms)\\
    \hline
    0 & 0.000(4) & 0.26(1) \\
    1 & 0.50(2)  & 0.32(4) \\
    2 & 0.810(7) & 0.23(2)\\
    3 & 1.222(9) & 0.39(4)\\
    4 & 1.664(7) & 0.26(2) \\
    \hline
    \end{tabular}
    \tablefoot{The first column gives the component number, the second the time of arrival in ms and the third the component width defined as the full width at half maximum (FWHM). Parenthesis give the one sigma uncertainties on the last digit.}
    \label{tab:FRB_components}
\end{table}

\subsection{Burst structure}
After dedispersing to the structure maximising DM, the pulse profile shows five distinct components with no visible scattering. In order to better characterise the intensity variation with time, we fitted the pulse profile to a five component gaussian and give the result in Table~\ref{tab:FRB_components}. The ToAs are given with respect to the arrival of the first component, and the component width is defined as the full width at half maximum (FWHM).
The resulting burst component widths are unresolved due to intra-channel dispersive smearing, so given our time resolution the scattering timescale must be negligible.
The total duration of the burst is 2.14(2)\,ms.

\begin{figure}
    \centering
    \includegraphics[width=\hsize]{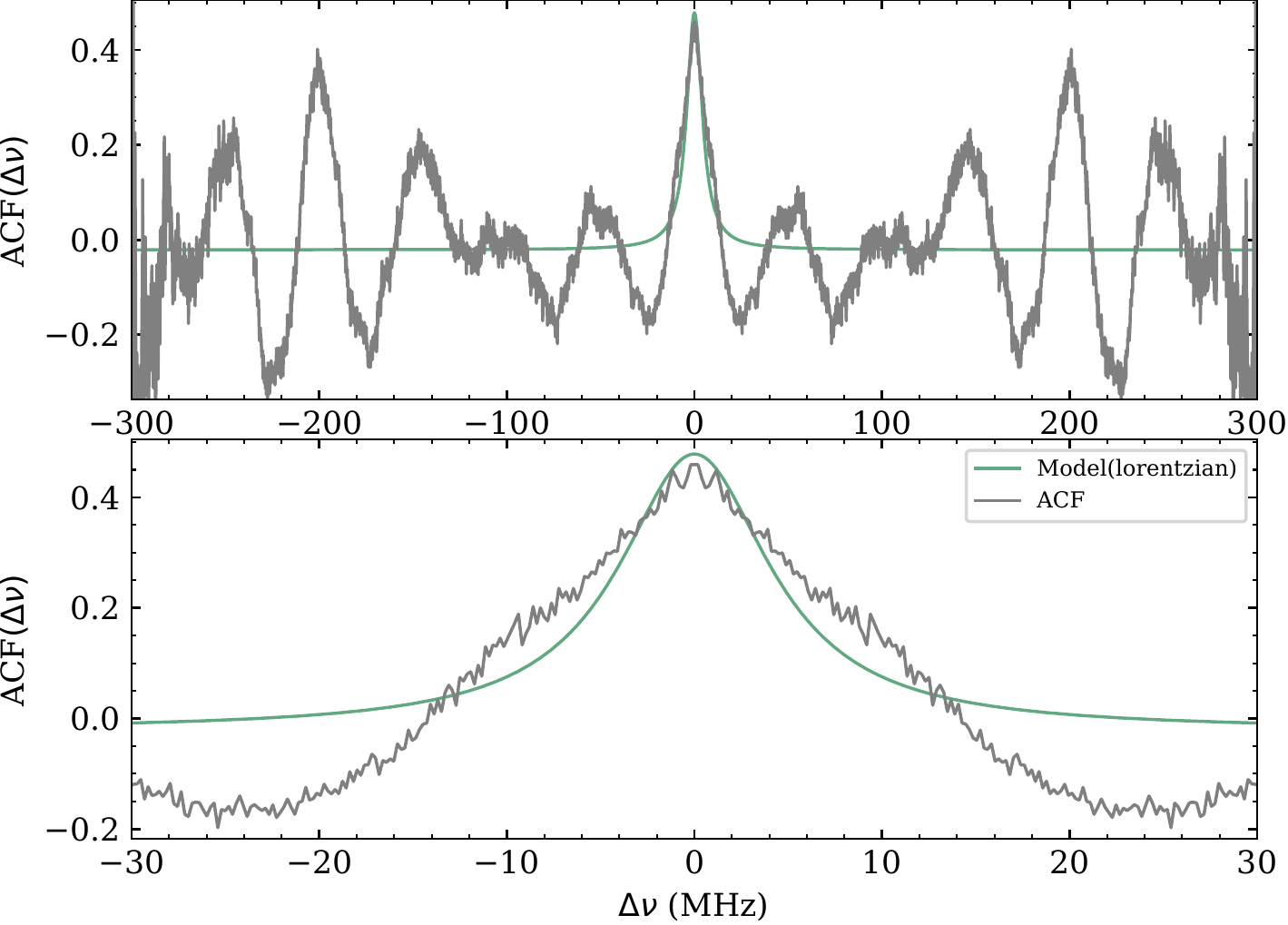}
    \caption{Auto-correlation function of the spectrum of \FRB. The grey line shows the ACF (see Eq.~\ref{eq:acf}) and the green line the lorentzian fitted to the central peak. The top panel shows the ACF of the whole observing bandwidth, while the bottom panel is a zoom in on the central peak.}
    \label{fig:FRB_acf}
\end{figure}

The burst shows frequency-dependent intensity variations, as expected from scintillation produced by propagation through the turbulent interstellar medium (ISM) \citep[see section 4.2.2 from][and references therein]{lorimer_handbook_2004}. In order to measure the scintillation bandwidth, we generated the auto-correlation function (ACF) of the burst averaged spectrum, defined as follows: 

\begin{equation}\label{eq:acf}
    \text{ACF}(\Delta\nu) = \dfrac{\displaystyle \sum_{\nu}(S(\nu))(S(\nu+\Delta\nu)) }{\displaystyle\sqrt{\sum_{\nu}(S(\nu))^2  \sum_{\nu}(S(\nu+\Delta\nu))^2}},
\end{equation}
where $S(\nu)$ is the burst averaged spectrum at frequency $\nu$ and $\Delta\nu$ the frequency lag. This is computed at the original frequency resolution.
%\ipm{Write formula?}\sms{I think that'd be good} 
After removing the zero-lag value, we fitted the central peak of the ACF to a lorentzian. The scintillation bandwidth is often defined as the full-width at half-maximum (FWHM) of the fitted lorentzian \citep{cordes_space_1986}. The YMW16 \citep{yao_new_2017} and NE2001 \citep{cordes_ne2001i_2003} models give the distribution of free electrons in the Milky Way (MW), and can thus be used to estimate the DM, scattering and scintillation contribution from the Galaxy.
Comparing the resulting scintillation bandwidth of $\Delta\nu_{\text{FRB}}=9.8(9)$\,MHz with the expected contribution from the MW in the direction of the FRB at 1370\,MHz, we find that it is consistent both with the YMW16 model, where ${\Delta\nu_{\text{YMW16}}\sim10.4}$\,MHz, and the NE2001 model, where ${\Delta\nu_{\text{NE2001}}\sim16.0}$\,MHz\footnote{Values estimated at 1370\,MHz with the \texttt{pygedm} package: \url{https://pygedm.readthedocs.io/en/latest/}}. The ACF and its fit to a lorentzian function are displayed in Fig.~\ref{fig:FRB_acf} %\ipm{ACF figure, fit to lorentzian, scintillation bandwidth and comparison to expected values from NE2001 and YMW16?}.

By eye, all burst components seem to cover a similar frequency extent. In order to thoroughly check whether the subcomponents exhibit a downward drift in frequency, we computed the 2D auto-correlation function of the burst, which we ultimately fitted to a 2D ellipse whose inclination gives a good estimate of the drift rate \citep{hessels_frb_2019}. The resulting 2D\,ACF, shown in Fig.~\ref{fig:acf2d}, shows an inclination angle consistent with being vertical, which means there is no subcomponent frequency drift. 

% python /home/arts/pastor/scripts/arts-analysis/scintillation_analysis.py -frb FRB201020 -o $outdir -s 4 -m cst

\subsection{Localisation}
\begin{figure}
    \centering
    \includegraphics[width=\hsize]{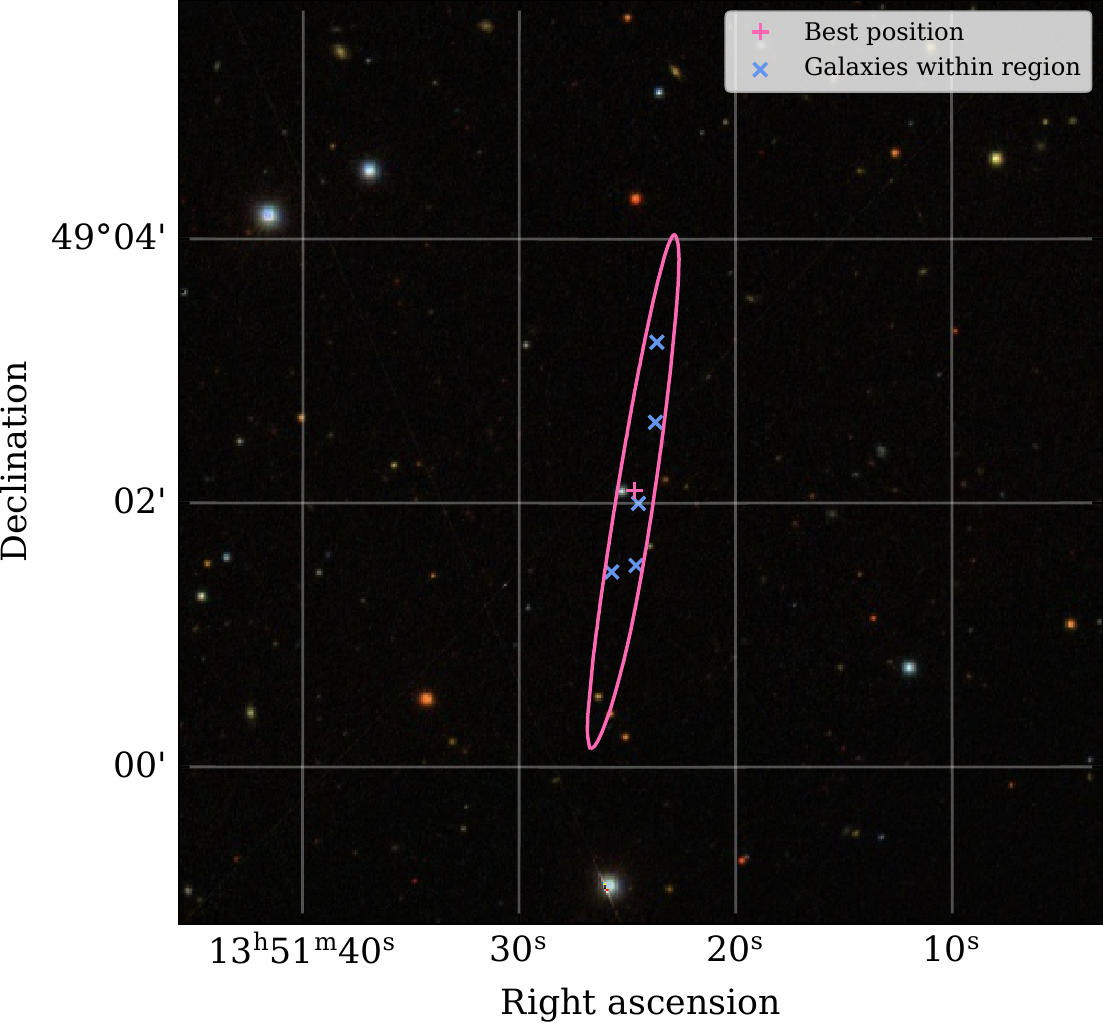}
    \caption{Localisation region of \FRB. The best position is indicated by a pink cross, and the 99\% confidence interval contour is represented by the pink ellipse. Galaxies from the NED database within the error region are marked as blue crosses. The background is a composite color image from the Sloan Digital Sky Survey \citep[SDSS,][]{blanton_sloan_2017}.}
    \label{fig:FRB_localisation}
\end{figure}

Since the WSRT is an array in the East-West direction, it can localise any detected FRB to a narrow ellipse \citep[cf.][]{connor_bright_2020}.
\FRB{} was detected in three adjacent CBs; CB29, CB35 and CB28. Following the localisation procedure as described in \cite{oostrum_fast_2020}\footnote{Localisation code can be found here: \url{https://loostrum.github.io/arts_localisation/}}, we used the SNR of the burst in all synthesised beams (SBs) of all CBs where it was detected to get its 99\% confidence region. We find the best position to be RA\,=\ra{13}{51}{25} and Dec\,=\dec{+49}{02}{06}. The localisation area is shown in Fig.~\ref{fig:FRB_localisation}, and it can be well described as $3.9\arcmin\times 17.2\arcsec$ ellipse with an angle of 103.9$\degr$ in the North-to-East direction.

The DM expected from the MW in the direction of the FRB is $\sim 22$\pccm\ according to the YMW16 model and $\sim29$\pccm\ according to NE2001. By assuming a MW halo contribution to the DM of $\sim50$\pccm\ \citep{prochaska_probing_2019}, we find an extragalactic DM of $\sim325$\pccm, that we use to estimate a redshift upper limit\footnote{The redshift upper limit was computed with the \texttt{fruitbat} package assuming a Planck18 cosmology and the Zhang 2018 method: \url{https://fruitbat.readthedocs.io/en/latest/}} of $z_{\text{max}}\sim0.43$ \citep{zhang_fast_2018, planck_collaboration_planck_2020, batten_fruitbat_2019}

We queried the NASA/IPAC extragalactic database (NED) to identify potential host galaxies of \FRB. We found five
galaxies within the error region catalogued in the SDSS-IV MaNGA \citep{graham_sdss-iv_2018}, which is based on the fourteenth data release of the sloan digital sky survey \citep[SDSS-DR14,][]{abolfathi_fourteenth_2018} and has a 95\% magnitude completeness limit of 22.2 in the \textit{g} filter\footnote{See SDSS-DR14 information here: \url{https://www.sdss.org/dr14/scope/}}. None of the five galaxies have a measured redshift or morphological type. Any of those five galaxies could thus be the host of \FRB, and there could be galaxies below the magnitude completeness of SDSS-DR14. The galaxies are listed in Table~\ref{tab:galaxies} of the Appendix~\ref{app:galaxies}.
%\ipm{Completeness level of the survey (see Emily's comment)}

% python /home/ines/Documents/projects/ARTS/scripts/localisation_hips.py -f FRB201020

\subsection{Polarisation and rotation measure} \label{sec:polcal}

The detection of \FRB{} triggered a dump of the full-Stokes data on CB29, allowing us to measure the polarisation properties of \FRB. We carried out calibration observations within 36\,h after the FRB detection. 
We pointed at the unpolarised source 3C147 and the linearly polarised source 3C286 to correct for the difference in gain amplitudes between the $x$ and $y$ feeds and the leakage of Stokes V into Stoke U, respectively \citep[cf.][]{connor_bright_2020}.
The polarisation properties of \FRB{} could provide us with essential information about the nature of this source. For instance, in the case of a rotating neutron star (NS), one would expect changes in the polarisation position angle (PPA) following the rotating vector model \citep[RVM,][]{radhakrishnan_magnetic_1969}. 

Unfortunately, the polarisation calibration observations were corrupted by radio frequency interferences (RFI) and their poor quality does not allow for robust PPA measurements.
Nevertheless, an estimate of the rotation measure (RM) can still be computed although with large uncertainties by assuming a constant value of Stokes V with frequency. From the resulting Stokes Q and U parameters, we obtain the best RM by applying RM synthesis \citep{burn_depolarization_1966, brentjens_faraday_2005}, and we check the resulting RM by applying a linear least squares fit to the position angle (PA) as a function of wavelength squared. We find an RM of $+110\pm69$\radsqm that can be visualised in Fig.~\ref{fig:rm} of the Appendix~\ref{app:rm}.

\subsection{Repetition rate}
The localisation region of \FRB{} has been observed for 107.8\,h with Apertif since the beginning of the survey in July 2019. This observing time includes follow up observations of both \FRB{} and FRB\,20190303A \citep{fonseca_nine_2020}. We thus set a 95\% upper-limit on the repetition rate of $\sim3.4\times10^{-2}\,\text{h}^{-1}$ assuming a Poissonian repetition process. This upper limit is roughly one order of magnitude lower than the average repetition rates observed in \RI{} and \RIII{} in observations carried with Apertif \citep{oostrum_fast_2020, pastor-marazuela_chromatic_2021}. However, we note this limit may be less constraining if the FRB has non-poissonian repetition as has been seen for other FRBs \citep{spitler_repeating_2016, oppermann_non-poissonian_2018, gourdji_sample_2019, the_chimefrb_collaboration_periodic_2020, hewitt_arecibo_2021}.
% \ipm{ Is this assuming Poisson repetition? Did you account for non-Poissonian repetition? What if the repetition of this FRB is highly non-Poissonian? and maybe cite some papers that indicate this is the case for many FRBs like Hewitt et al. or the CHIME/FRB 16 day periodicity paper.}
% \sms{Will you add a sentence to conclude the rate? Now only time on sky at those coordinates is given.}
%-------------------------------------------------------------
% \section{Results}

\section{Timing analysis}
\label{sec:results}

We applied several timing techniques in order to determine the presence of a periodicity in \FRB, and the \RIII\ bursts A17 and A53. 
Initially, we obtained the power spectra of each pulse profile, defined as the Leahy normalised \citep{leahy_searches_1983} absolute square of the fast Fourier transform (FFT), to identify any potential peaks in the power spectrum of each burst. 

Additionally, all three pulse profiles were fitted to multi-component gaussians in order to determine the time of arrival (ToA) of each burst subcomponent. The ToAs as a function of component number were subsequently fitted to a linear function using a weighted least squares minimization in order to determine the mean subcomponent separation \Psc{} as well as the goodness of fit using two different statistics. We used the statistic $\hat{S}$ as described in \cite{the_chimefrb_collaboration_sub-second_2021} in order to have a direct comparison between our sample and the FRBs presented there. 
Furthermore, we used the reduced $\chi^2$ (\rchisq) statistic in order to take into account the statistical error on the ToAs. We also applied the \rchisq{} statistic to \chimefrbs{} for comparison. Given the necessity of fitting a model to the tail of the distribution of the simulated statistics for \FRBperiodic{} in \citep{the_chimefrb_collaboration_sub-second_2021} and the high significance found, we do not compare the statistics of \FRBperiodic{} with \FRB{} in this work. The \rchisq{} statistic seems to be more adapted to the ToAs and errors obtained through our pulse profile fitting routine than the $\hat{S}$ statistic, and it does not significantly alter the periodicity significance of the \chimefrbs. We thus find it to be more robust when computing periodicity significances. %\sms{added a slash to create a space after the chime frbs}

\begin{figure*}
    \centering
    \includegraphics[width=17cm]{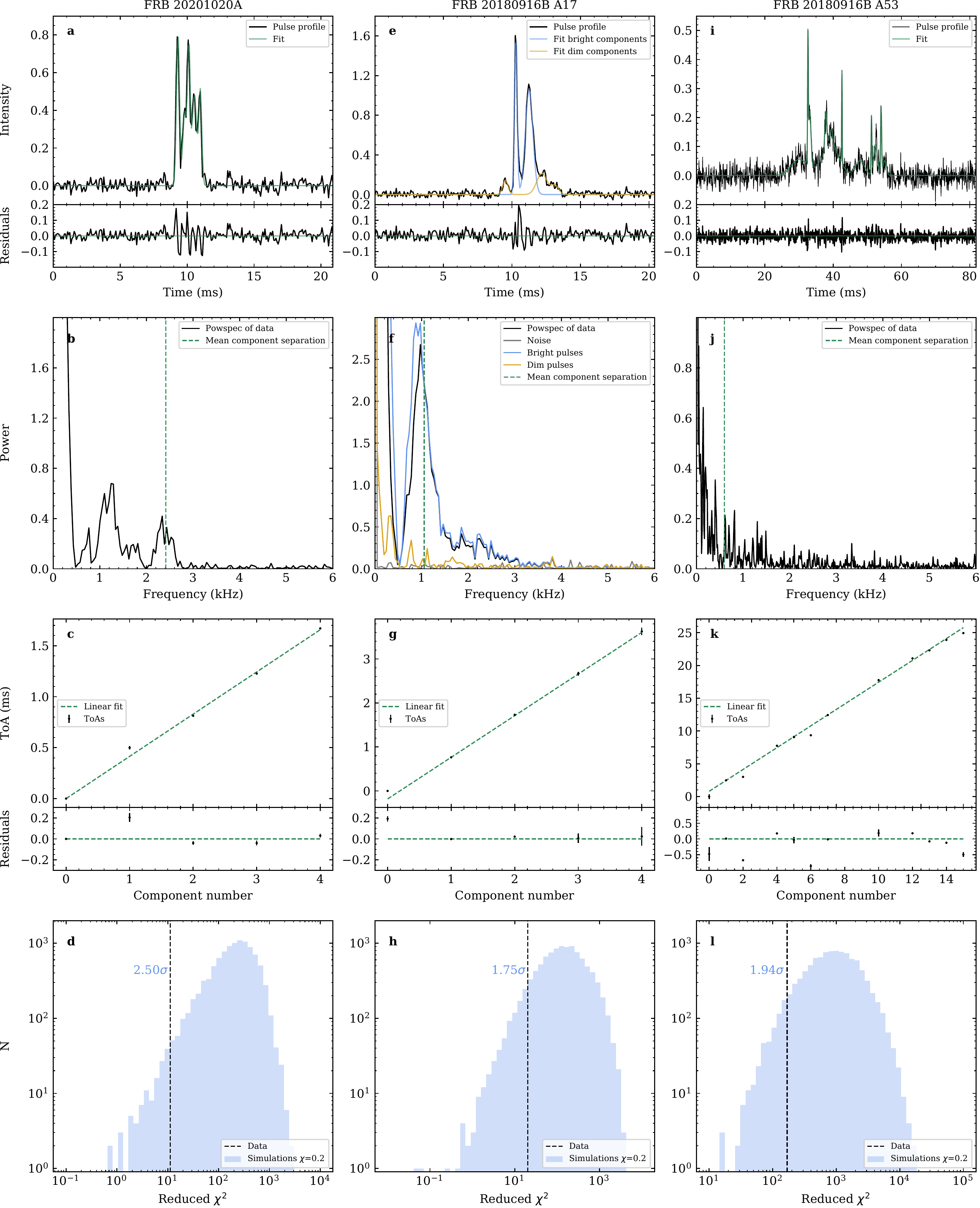}
    \caption{Timing analysis of \FRB{} (left column), A17 (central column) and A53 (right column). The top of panels a, e and i show the pulse profile of each burst (black) and the fitted multi-component gaussians in green for \FRB{} and A53; for A17 the bright components are shown in blue and the dim components in yellow. The bottom of panels a, e and i shows the residuals of the multi-component gaussian fit of each pulse profile. Panels b, f and j show the power spectra of each pulse profile in black. In b, the vertical dashed line indicates the average separation of the \FRB{} components. In f, the blue and yellow lines show respectively the power spectra of the bright and dim components. Panels c, g, and k show the ToAs of all subcomponents as a function of component number. The dashed green line in c, g, and k are linear fits to the ToAs, and the lower panels show their residuals. Panels d, h, and l show histograms of the simulated \rchisq{} statistic to compute the significance of the periodicity. The vertical lines correspond to the \rchisq{} of the linear fit to the data. The significance is indicated as blue text.}
    \label{fig:periodicity_analysis}
\end{figure*}

Lastly, we computed the significance of the potential periodicities by simulating the ToAs of $10^5$ bursts following the procedure presented in \cite{the_chimefrb_collaboration_sub-second_2021}.
Figure~\ref{fig:periodicity_analysis} contains the relevant plots resulting from the timing analysis, and Table~\ref{tab:periodicity_significance} gives the significance of the FRB periodicities.

% Using Stingray to compute power spectrum. Stingray is built to work with counts, so values always >= 0. Otherwise unable to compute noise power spectrum (~half of the values are negative).

\begin{table*}
\caption{Statistical significance of the FRB periodicities}             
\label{tab:periodicity_significance}      
\centering          
\begin{tabular}{l l l l l l l}
\hline\hline       
FRB & $n_{\text{comp}}$ & \Psc{} (ms) & \phantom{00}\rchisq & \schisq & \phantom{0}$\hat{S}$ & \sshat \\
\hline                    
\FRB & (0,1,2,3,4) & \phantom{0}0.415(6) & \phantom{00}11.17 & 2.50 & \phantom{0}4.18 & 1.41 \\
\RIII{} A17 & (0,1,2,3,4) & \phantom{0}0.95(3) & \phantom{00}20.33 & 1.75 & 10.47 & 3.54\\
\RIII{} A53 & (0,1,2,4,5,6,7,10,12,13,14,15) & \phantom{0}1.7(1) & \phantom{0}168.3 & 1.94 & \phantom{0}5.12 & 0.95\\
\hline
\FRBfivecomp & (0,1,2,3,5) & \phantom{0}2.8 & 2027.41 & 1.90 & \phantom{0}5.13 & 1.89 \\
\FRBsixcomp & (0,1,2,3,4,5) & 10.8 & \phantom{0}270.80 & 2.96 & \phantom{0}7.42 & 2.56 \\
\hline                  
\end{tabular}
\tablefoot{Values computed for the FRBs presented in this paper as well as the FRBs below the 3$\sigma$ significance threshold from \cite{the_chimefrb_collaboration_sub-second_2021}. For each FRB considered, we give the component number vector ($n_{\text{comp}}$), the quasi-period \Psc{} in milliseconds, the reduced $\chi^2$ (\rchisq) with its respective significance \schisq, and the value of the statistic $\hat{S}$ with its respective significance \sshat{} as recalculated in this work.}
\end{table*}

% Version with original formatting

%\begin{table*}
%\caption{Statistical significance of the FRB periodicities}             
%\label{tab:periodicity_significance}      
%\centering          
%\begin{tabular}{c c c c c c c}
%\hline\hline       
%FRB & $n_{\text{comp}}$ & \Psc{} (ms) & \rchisq & \schisq & $\hat{S}$ & \sshat \\
%\hline                    
%\FRB & (0,1,2,3,4) & 0.415(6) & 11.17 & 2.50 & 4.18 & 1.41 \\
%\RIII{} A17 & (0,1,2,3,4) & 0.95(3) & 20.33 & 1.75 & 10.47 & 3.54\\
%\RIII{} A53 & (0,1,2,4,5,6,7,10,12,13,14,15) & 1.7(1) & 168.3 & 1.94 & 5.12 & 0.95\\
%\hline
%\FRBfivecomp & (0,1,2,3,5) & 2.8 & 2027.41 & 1.90 & 5.13 & 1.89 \\
%\FRBsixcomp & (0,1,2,3,4,5) & 10.8 & 270.80 & 2.96 & 7.42 & 2.56 \\
%\hline                  
%\end{tabular}
%\tablefoot{Values computed for the FRBs presented in this paper as well as the FRBs below the 3$\sigma$ significance threshold from \cite{the_chimefrb_collaboration_sub-second_2021}. For each FRB considered, we give the component number vector ($n_{\text{comp}}$), the quasi-period \Psc{} in milliseconds, the reduced $\chi^2$ (\rchisq) with its respective significance \schisq, and the value of the statistic $\hat{S}$ with its respective significance \sshat{} value as recalculated in this work.}
%\end{table*}

% jupyter notebook /home/ines/Documents/projects/subpulse_periodicity/scripts/subpulse_periodicity_plots.ipynb

\subsection{\FRB}
\label{sec:TOAs}
To get the mean subcomponent separation \Psc, we fitted the ToAs given in Table~\ref{tab:FRB_components} as a function of component number $n = (0,1,2,3,4)$. We get \Psc$=0.415(6)$\,ms, which we assume to be the period whose significance we want to examine.

The power spectrum of the FRB shows a peak at the frequency corresponding to the mean subcomponent separation, 2409\,Hz. 
% Adding this
However, it also displays a peak with a higher amplitude at half the frequency corresponding to \Psc. 
We argue that this peak is more prominent because of the higher amplitude of components 0, 2 and 4, which are spaced by twice the \Psc. 
Additionally, the ToA of component 1 is delayed with respect to the others, as can be seen in Fig.~\ref{fig:periodicity_analysis} panel c, thus lowering the power of the peak at 2409\,Hz. Red noise could also play a factor in increasing the peak amplitude. We thus consider 2409\,Hz to be the fundamental frequency, and interpret the 1200\,Hz peak as a lower-frequency harmonic.

By using the \rchisq{} as the reference statistic and applying it both to the data and the simulations, we obtain a significance of $2.50\sigma$. The different steps of the analysis are plotted in Fig.~\ref{fig:periodicity_analysis}, panels a-d. 
While \schisq{} and \sshat{} are not significant enough for a conclusive periodicity detection, this FRB is visually analogous the \chimefrbs. This suggests that these three bursts belong to the same morphological class of FRBs that present quasi-periodic components in time, with all subcomponents showing a similar frequency extent.
%Although both \schisq{} and \sshat{} are below the $3\sigma$ threshold, both these values are comparable to what is obtained for \chimefrbs. This suggests that these three bursts belong to the same morphological class of FRBs that present quasi-periodic components in time, with all subcomponents peaking at the same frequency.

\subsection{\RIII{} A17}
%\sms{perhaps add FRB20180916B to the subsection title such that it's clear without reading the text that these are bursts from R3}
The pulse profile of A17 can be well fitted by a five component gaussian, with the second and third components being much brighter than the others. 
Its power spectrum shows a bright peak at $\sim$1000\,Hz. Since the separation between the second and third components is $\sim$1\,ms, we tested whether the power spectrum was dominated by these two bright subcomponents instead of arising from an intrinsic periodicity. 
To do so, we created fake pulse profiles of the "bright" and "dim" components.
The pulse profile of the bright components was created by subtracting the gaussian fit of the first, fourth and fifth components from the data, and the pulse profile of the dim components by subtracting the fit of the second and third components from the data.
Next we generated power spectra of the "bright" and "dim" pulse profiles, both over-plotted in Fig.~\ref{fig:periodicity_analysis}, panel f. Since the power spectrum of the bright components largely overlaps with the power spectrum of the original pulse profile, we conclude that the peak in the power spectrum is dominated by the two brightest components.

We further fitted the ToAs to a linear function and compared the resulting \rchisq{} and $\hat{S}$ statistics to simulations. While \schisq$=1.75\sigma$ is below the $3\sigma$ threshold, \sshat$=3.54\sigma$ is not. Due to this discrepancy, we cannot confirm nor rule out the presence of a periodicity in A17. The steps of the timing analysis are plotted in Fig.~\ref{fig:periodicity_analysis}, panels e-h.

\subsection{\RIII{} A53}

A53 is the widest \RIII{} burst and with the highest number of components ($\geq11$) presented in \cite{pastor-marazuela_chromatic_2021}, making it a good candidate for periodicity analysis.
Although the ToAs plotted as a function of component number show irregularities in the spacing between bursts, we still carried out the fit to a linear function and the significance computation by associating the larger "gaps" to missing components, resulting in a component number vector $n=(0,1,2,4,5,6,7,10,12,13,14,15)$. 
We find \schisq$=1.94\sigma$ and \sshat$=0.95\sigma$. 

The power spectrum of A53 is well fitted by a power law and all fluctuations are consistent with noise. We do not find a significant peak in the power spectrum at the average component separation of 1.7\,ms.
We thus conclude that A53 does not show a significant periodicity in its structure. All steps of the analysis are shown in Fig.~\ref{fig:periodicity_analysis}, panels i-k.

%-------------------------------------------------------------
\section{Discussion}
\label{sec:discussion}

We next consider which mechanism can best explain the  periodicity in \FRB: 
the rotation of an underlying \mbox{sub-millisecond} pulsar (\S \ref{sec:smsp}), 
the final orbits of a compact object merger (\S \ref{sec:bns}), 
crustal oscillations after an X-ray burst (\S \ref{sec:osc}),
or
equidistant emitting regions on a rotating neutron star (\S \ref{sec:sparks}).
We also comment on whether these FRBs represent a new morphological type.

\subsection{Sub-millisecond pulsar}
\label{sec:smsp}
One of the potential origins of the periodic structures in \FRB{} and the \chimefrbs{} discussed in
\cite{the_chimefrb_collaboration_sub-second_2021} is the rotation of a neutron star with beamed emission, analogous to Galactic radio pulsars. While the period range observed in the aforementioned FRBs is compatible with Galactic pulsars,
the quasi-period of \FRB{} \Psc$\sim$0.415\,ms is in the sub-millisecond regime.

A spin rate of such magnitude is interesting because it could provide the energy required for the highly luminous radio emission seen in FRBs in general. 
For a rotation-powered neutron star, the spin rate $\nu$ and surface magnetic field strength $B_\mathrm{surf}$ determine the potential of the region that accelerates the particles that generate the radio bursts.
% add \citet{rs75} sentence? 
For example, giant pulses are only common in those pulsars with the highest known values of $\nu^3B_\mathrm{surf}$, interpreted as the magnetic field strength at the light cylinder $B_\mathrm{LC}$ \citep[where $B_\mathrm{LC}>2.5\times10^{5}$\,G,][]{johnston_giant_2004,Knight2006}. Clearly the spin frequency is an important driver powering such luminous emission. 

% Could sub-ms sources underlie all FRBs? That appears inconsistent with the observations of FRBs that last for over 5\,ms. But these bursts could well consist of subcomponents that are blended together due to effects of the intervening medium and observational limitations. 
Fast radio bursts lasting for over 5\,ms seem to be at odds with a sub-millisecond NS interpretation.
However, these bursts could consist of subcomponents that are blended together due to scattering or observational limitations such as temporal resolution and intra-channel smearing.
In the general population, the intrinsic width of the emitted FRBs is not well constrained \citep{gardenier_multi-dimensional_2021}.
A rotating millisecond neutron star or magnetar may thus explain the bright emission of an FRB, even when only a single burst is observed. Although the component separation in \FRBperiodic{} is two orders of magnitude larger than in \FRB, the number of FRBs with regularly spaced sub-second components is so small at the moment that one should aim for a single explanation.
Overall, a sub-millisecond neutron star is a highly interesting hypothesis to test.

The  \FRB{}  quasi-period is
shorter than that of the fastest spinning pulsar confirmed so far, with $\sim$1.4\,ms \citep{hessels_radio_2006},
and would be the smallest known neutron-star spin period.
 The maximum rotation rate an NS can achieve can set important constraints on the NS equation of state (EoS) \citep[e.g.][]{shapiro_implications_1983}. This maximum rotation rate is given by the mass-shedding frequency limit, above which matter from the outer layers of the NS is no longer gravitationally bound and is thus ejected. \cite{haensel_keplerian_2009} establish an empirical formula for the mass-shedding frequency $f_{\text{max}}$ given by the mass and radius of the NS, with $C\sim1$\,kHz,
%No sub-millisecond pulsars have been discovered so far, and their existence has been discussed for decades. 
\begin{equation}\label{eq:fmax}
    f_{\text{max}} (M) = C \left( \frac{M}{M_{\odot}} \right)^{1/2}\left(\frac{R}{10\text{km}}\right)^{-3/2}.
\end{equation}

\begin{figure}
    \centering
    \includegraphics[width=\hsize]{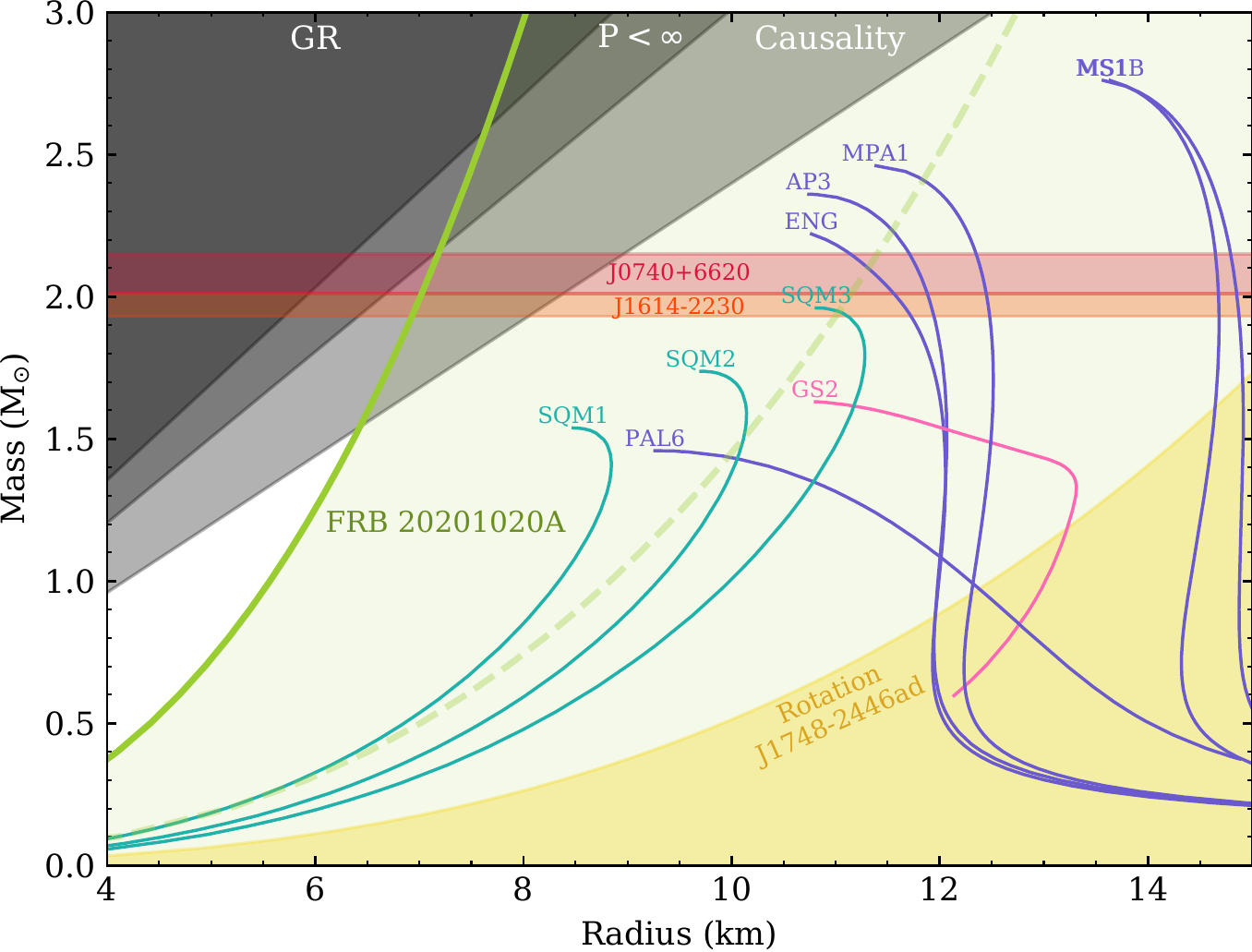}
    \caption{Neutron star mass-radius diagram. The solid lines represent the mass-radius relationship of typical EoSs. The gray shaded regions are forbidden by the general relativity (GR), finite pressure (P$<\inf$) and Causality. The yellow shaded region is forbidden by the rotation of the fastest spinning pulsar J1748$-$2446ad
    \citep{hessels_radio_2006}. The red shaded region shows the mass of J0740+6620 \citep{cromartie_relativistic_2020,fonseca_refined_2021} and the orange region the mass of J1614-2230 \citep{demorest_two-solar-mass_2010}, the two heaviest known NSs. Valid EoSs need to cross those regions. The green line and green shaded region would be forbidden if the quasi-period of \FRB{} was produced by the rotation of a NS at 2409 Hz, and the dashed green line if the rotation was at 1205 Hz.}
    \label{fig:ns_eos}
\end{figure}

Figure~\ref{fig:ns_eos} shows a mass-radius diagram based on \cite{demorest_two-solar-mass_2010}, including typical EoSs, the physical limits set by causality, finite pressure and general relativity \citep{lattimer_nuclear_2012}, and observational constraints given by the two most massive known NSs \citep{cromartie_relativistic_2020, fonseca_refined_2021, demorest_two-solar-mass_2010} and the fastest spinning NS at 716\,Hz \citep{hessels_radio_2006}. 
In order to check whether the quasi-period of \FRB{} could be explained by the rotation of a NS, we plot the mass-radius relation using Eq.~\ref{eq:fmax} assuming the mass-shedding frequency to be given by \FRB, $f_{\text{max}}=1/0.415$\,ms$\sim2409$\,Hz. Since no EoSs remain in the white region of the diagram, we conclude that the quasi-period of \FRB{} is incompatible with being due to the rotation of a NS using typical EoSs. %We also consider the case where $f_{\text{max}}=1/0.83$\,ms$\sim1205$\,Hz, or twice the subcomponent separation (see peak in power spectrum in Fig.~\ref{fig:periodicity_analysis}b). While some EoSs cross to the white region of the diagram, they would be at odds with the observed NS masses.

If, given the similar morphological properties of \FRB{} with the CHIME/FRB FRBs \FRBperiodic, \FRBsixcomp{} and \FRBfivecomp, the four
FRBs are originated by the same type of progenitors,
the periodicity of the burst components is unlikely to be due to the rotation of a NS.

% python /home/ines/Documents/projects/subpulse_periodicity/scripts/ns_eos.py

\subsection{Compact object merger}
\label{sec:bns}

Merging compact objects have been hypothesised to produce FRBs through magnetic interaction between the two bodies in the system, including binary neutron star systems \citep[BNS,][]{piro_magnetic_2012, lyutikov_electromagnetic_2013, totani_cosmological_2013, wang_fast_2016, hansen_radio_2001}, black hole-neutron star systems \citep[BHNS,][]{mcwilliams_electromagnetic_2011, mingarelli_fast_2015, dorazio_bright_2016}, and white dwarfs (WD) with either a neutron star or a black hole \citep{liu_model_2018, li_model_2018}.
% \ipm{model references} 
The presence of multiple peaks in the FRB pulse profile could be explained by a magnetic outburst extending through successive orbits of the binary system, and thus the subcomponent frequency should match half the gravitational wave (GW) frequency if one object in the system produces bursts, or the GW frequency if both of them do.
The maximal GW frequency is attained when the binary reaches the mass-shedding limit for an NS \citep{radice_dynamics_2020}. This frequency is given by the EoS of the merging NS, which is expected to be between 400 and 2000\,Hz \citep{bejger_impact_2005,dietrich_interpreting_2021,hanna_method_2009}. The frequency seen in the components of \FRB{} of 2409\,Hz is above the typical limits and it could only be explained by a very soft EoS.

This FRB frequency could only be produced at the very last moments of the inspiral, right before the merger. At this stage, the frequency derivative would be very high, which translates as a perceptibly decreasing spacing between burst subcomponents. 
Following \citep[][equation 16]{cutler_gravitational_1994}, we can derive the analytical expression for the phase of the gravitational waveform as a function of time, $\phi(t)$, by solving the differential equation of $\frac{df}{dt}$ and integrating $f(t)=\frac{d\phi}{dt}(t)$, where $f(t)$ is the GW frequency. We compute the maximal number of orbits that could be completed between the time when the system reaches the FRB frequency and the time of the merger through the maximal phase difference $\Delta\phi_{\text{max}}$, where one orbit equals $2\pi$\,rad, with the steps of the computation detailed in Appendix~\ref{app:dphi_max}. We find the following expression:

\begin{equation} \label{eq:dphi_max}
    \Delta\phi_{\text{max}} = \frac{1}{16\pi^{5/3}}\left(\dfrac{G\mathcal{M}}{c^3}\right)^{-5/3} f_{\text{FRB}}^{-5/3}
\end{equation}

Here, $\mathcal{M}\equiv\mu^{3/5}M^{2/5}$ is the chirp mass, with $M=M_1+M_2$ the total mass of the system and $\mu=M_1M_2/M$ the reduced mass. $G$ is the gravitational constant and $c$ the speed of light in the vacuum. 
We have computed $\Delta\phi_{\text{max}}$ for three values of $M_1=0.2, 1.4, 100$\,M$_{\odot}$, and a range of $M_2$ values between 0.2\,M$_\odot$ \citep[close to the lowest mass WD;][]{kilic_lowest_2007} and 100\,M$_\odot$. We have tested both $f_0=$2409\,Hz and 4818\,Hz as initial GW frequencies.

\begin{figure}
    \centering
    \includegraphics[width=\hsize]{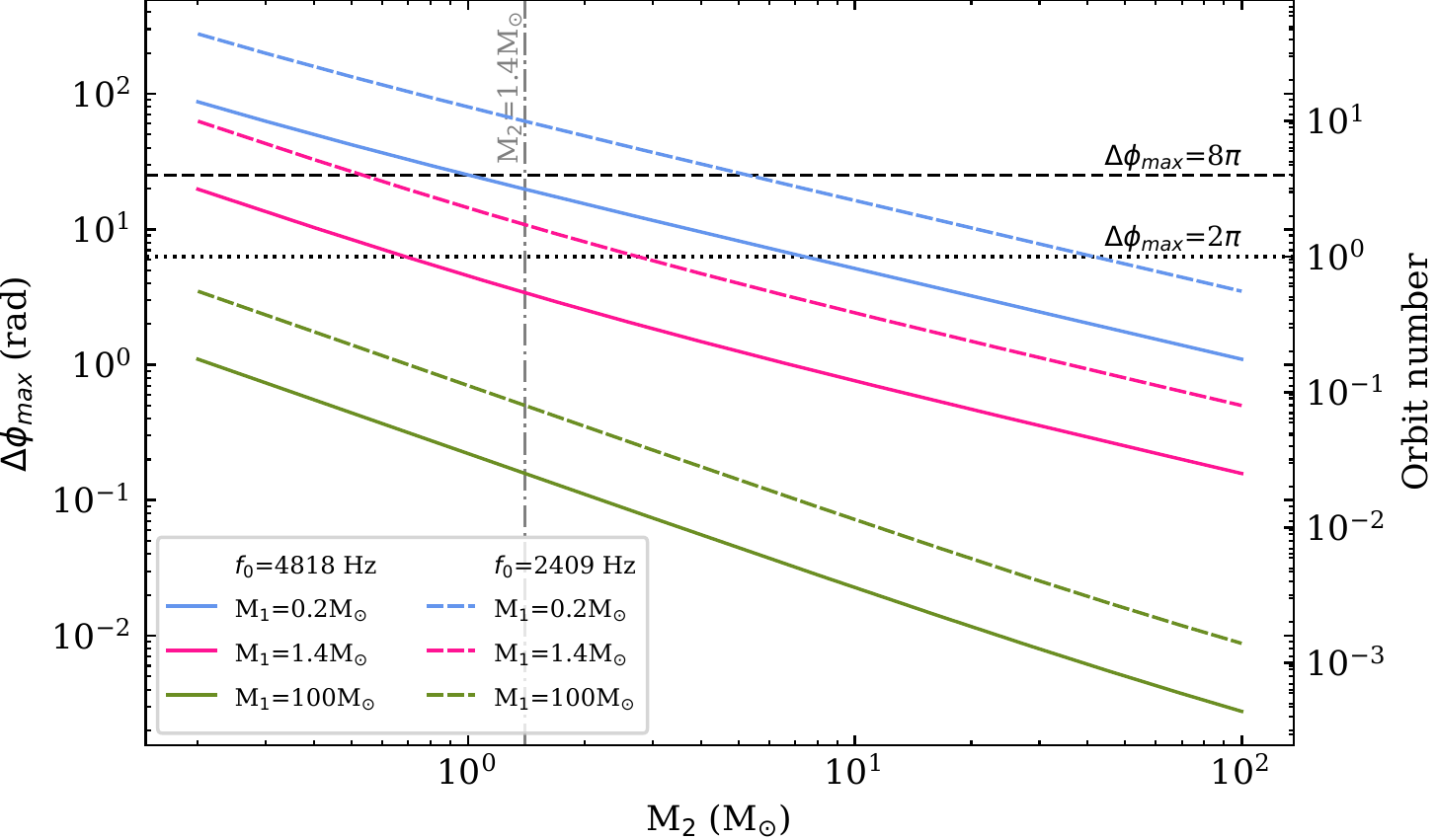}
    \caption{Maximal orbital phases a binary system can complete before a merger after reaching a GW frequency of 2409\,Hz (solid lines) and 4818\,Hz (dashed lines). The dashed and dotted black horizontal lines mark respectively phases of $8\pi$ (four rotations) and $2\pi$ (one rotation), required to see five and three FRB subcomponents respectively. The orbital phases have been computed for binary systems with the mass of the first compact object being M$_1$=0.2\,M$_{\odot}$ (blue), 1.4\,M$_{\odot}$ (pink) and 100\,M$_{\odot}$ (green), while the mass of the second object M$_2$ ranges between 0.2 and 100\,M$_{\odot}$. The dot-dashed gray vertical line indicates M$_2=1.4$\,M$_{\odot}$.}
    \label{fig:CO_mergers}
\end{figure}

As shown in Fig.~\ref{fig:CO_mergers}, only a WD-WD system could complete four rotations ($\Delta\phi_{\text{max}}=8\pi$) after reaching $f_{\text{FRB}}=4818$\,Hz. However, given the typical WD radius of 5000-10000\,km, this is not a realistic scenario. Indeed, we can get an estimate of the distance between the two compact objects in the system when they reach the FRB frequency using Kepler's third law, $r=\frac{(GM)^{1/3}}{(\pi f)^{2/3}}$. For a binary WD system of $\sim1$\,M$_\odot$ each, the expected separation would be $\sim16$\,km, well below the typical WD radius. Since none of the systems with masses corresponding to a BNS or BHNS system could reach $\Delta\phi_{\text{max}}=8\pi$, we can discard the pre-merger scenario as an explanation for \FRB. 

%Following \citep[][equation 16]{cutler_last_1993} and assuming two compact objects with a mass of 1.4\,M$_\odot$, we find the frequency derivative to be $\dot{f}\sim2\times10^6$\,s$^{-2}$ for an orbital frequency of 2409\,Hz, which would translate as a separation of $\sim0.3$\,ms. The separation between components 3 and 4 would be similar to the instrument resolution. Besides, when integrating $\dot{f}(t)$ twice to get the orbital phase $\phi(t)$, we find that the binary system cannot complete a full orbit after reaching a frequency of 2409\,Hz, making the detection of multiple bursts impossible. We thus rule out the pre-merger scenario as a potential progenitor for \FRB.

The quasi-period of \FRBfivecomp{} is also challenging to explain within this model \citep{the_chimefrb_collaboration_sub-second_2021}. Since only a small fraction of the known FRBs show regularly spaced components, we argue that these belong to the same class of progenitor and hence disfavour compact object merger models as the progenitors of FRBs with this type of morphology.

% ~/Documents/projects/subpulse_periodicity/subpulse_periodicity.ipynb

% \ipm{Formation of an unstable, rapidly spinning magnetar right after BNS merger that emits radio waves right before collapsing into a BH?}

\subsection{Crustal oscillations}
\label{sec:osc}

If FRBs are a result of short X-ray bursts from a magnetar, the quasi-periodicity we observe could be explained through crustal oscillations of said neutron star. These oscillations would result from the same X-ray burst that also caused the FRB~\citep{wadiasingh_fast_2020}.
The sub-bursts of \FRB{} suggest an oscillation  frequency of $\sim$2400~Hz.
~\cite{wadiasingh_fast_2020} examine if FRB sub-pulses can be a result of torsional crustal oscillations during a short X-ray burst. 
If so, their simulations show that the fundamental eigenmodes with low multipole number (corresponding to frequencies between 20--40~Hz) would be more prevalent than the higher frequencies that we observe. We conclude the difference in frequencies is too large, and disfavour crustal oscillations as the source of the periodicity. 

%\subsection{Distinct sparks of emission in NS magnetospheres} \ipm{Title?}
\subsection{Ordered emission regions on a rotating magnetar}
\label{sec:sparks}

% We have ruled out the sub-millisecond pulsar and the compact object merger scenarios as progenitors of \FRB. 

The burst morphology and time scales of \FRB{} are very similar to two types of quasi-periodic behaviour seen in regular pulsars:
subpulses and microstructure.
Below we discuss scenarios where the FRB is produced in a similar manner, but on a magnetar instead of a purely rotation-powered pulsar. 
%\sms{Could add "rotation-powered" before pulsar to make distinction even more clear} DONE-JVL

\subsubsection{The analogue of pulsar subpulses}

%We have ruled out the sub-millisecond pulsar and the compact object merger scenarios as progenitors of \FRB. 

%% We now investigate a scenario where the bursts originate from the magnetosphere of a rotating magnetar whereby the burst morphology reflects 
%% its intrinsic substructure, caused by distinct emission regions that move through our line of sight as a star rotates. 
 %\ab{I am not sure about semantics here: magnetars are also neutron stars, right? Or it is customary to separate rotationally and magnetically powered NS?}

Like the components in \FRB{}, the individual pulses of rotation-powered pulsars are usually composed of one or more sub-pulses. Subpulse widths lie in the range of $1\degree-10\degree$ of the spin longitude \citep[Table 16.1 in][]{Lyne2012}, thus broadly spanning $0.3-140$\,ms for pulsars with spin periods of $0.1-5$\,s. This qualitatively fits the individual components widths of observed FRB structures.

For a substantial fraction of these pulsars (mostly those with longer periods) the position or intensity of these subpulses 
changes in an organized fashion from one spin period to another \citep{Weltevrede2007,Basu2019}. 
The most widely accepted theory of drifting subpulses interprets these as emission
from distinct plasma columns originating in a rotating carousel of discharges within the polar gap \citep{1975ApJ...196...51R}. 
As the neutron star spins, these ordered, regularly spaced emission regions ("sparks") produce quasi-periodic 
bursts in our line of sight.
In this case, the burst morphology directly reflects the spatial substructure of the distinct emission regions.
%The alternative theory explains drifting subpulses as a product of modulation caused by global oscillation modes of
%the neutron star \citep{Rosen2008}. 
Given the similarity in appearance, we hypothesize here that a magnetar produces the components in \FRB{} 
in the same way that a pulsar produces subpulses.

This carousel rotation causes  subpulses to appear at progressively changing longitudes\footnote{There is also non-drifting, intensity-modulated periodicity, but that requires a spin period smaller than the observed periodicity, which we have ruled out in \S \ref{sec:smsp}.}. 
%, corresponding to grazing (close to the cone edge) traverse of the line of sight through emission cone. 
Every period may contain one or few individual subpulses roughly equidistant from each other. 
These regular subpulses cause a stable second periodicity on top of the primary, rotational periodicity.
In the most regular cases, the resulting drifting subcomponents behave quite periodically; 
in PSR~B0809+74, the textbook example \citep{van_leeuwen_probing_2003}, 
the spark rotation and spacing were recognised early on to be steady to over 1 part in 250 \citep{unwin_phase_1978}.
Overall, the spark regularity is sufficient to produce the quasi-periodicity we observe here. 

The residuals in the timing fit of \FRB{} (Sect.~\ref{sec:TOAs}) have an analogue in pulsar sub-pulses, too.
There, the deviation from strict periodicity within a single pulsar rotation is well-documented for some of the sources with more stable drift and is well described by the curvature of the line-of-sight path through the emission cone \citep{Edwards2002, 2002A&A...387..169V}. At least for some sources the drifting rate can be measured with good precision (indicating prominent periodic modulation) and it has been shown that it can slowly vary with time \citep{Rankin2006}, jitter on small timescales \citep{van_leeuwen_plasma_2012}, and change rapidly during a mode switch \citep{Janagal2022}. Several drift periodicities can even be present at the same time at different spin longitudes \citep[bi-drifting,][]{Szary2020}.

% \ab{I don't know whether P2 can evolve with time / mode switch, Joeri?}

Pulsar subpulses are not only similar to the components in \FRB{}, the \chimefrbs{} and A17
with respect to their regular spacing; also the number of sub-bursts agrees.
Quite a few pulsars, such as 
B0329+54 \citep{mitra_absolute_2007} 
% \ab{this paper is about polarization, not drift. Am I missing something?} \ipm{it is about polarization but it shows multi-component pulse profiles},
% B0826$-$34 \citep{van_leeuwen_plasma_2012},
B1237+25 \citep{srostlik_core_2005},
% J1740+1311 \citep{force_subpulse_2010},
% B1831$-$04 \citep{lorimer_binary_2008}, 
and B1857$-$26 \citep{mitra_subpulse_2008},
show both a similarly high number of distinct subpulses and quasi-periodically spaced single pulse profiles. 
% \ab{I would expand more on why you are citing this papers}.

One concern in this analogy  might be the timescale of the ``on-mode'' for FRBs. 
If the FRBs we discuss here come from a single neutron-star rotation, the absence of fainter bursts nearby means 
the  subpulses pattern flares for one period, and then turns off. This may seem problematic.
Yet, for pulsars that display nulling, or different  modes of emission, the spark pattern is known to establish itself very quickly, at the timescale of less than a spin period \citep{Bartel1982}.
Furthermore, pulsars such as J0139+3336 \citep[a 1.25\,s pulsar that emits a single pulse only once every $\sim$5 minutes;][]{Michilli_single_2020} and   
the generally even less active RRATS \citep{McLaughlin_rrats_2006} 
show that a population of only very sporadically, but brightly, emitting neutron stars exists. %\ipm{What about the upper limit on the repetition rate?}

% and sources such as 
% however to our best knowledge no records have been made of emission disappearing after one rotation. \ab{I'm not sure about this paragraph,  since I do not actually know the flux limits on potential adjacent FRBs. Perhaps the sensitivity is not enough to catch them assuming standard pulsar pulse-to-pulse variability rate.}

Overall, while 
the amount of information available for FRBs is limited, their behaviour falls within the (admittedly broad)
range of features exhibited by subpulse drift.
% , it is so far hard to judge whether drifting subpulses can indeed be the cause of the observed periodicity. 

\subsubsection{The analogue of pulsar/magnetar microstructure}

% PERIODICITY + RELATIONSHIP
In pulsars, the individual subpulses mentioned in the previous sections 
themselves often contain yet another level of $\mu$s$-$ms,
substructure \citep{kramer_high-resolution_2002}. 
Often narrow micropulses are observed in 
groups of several spikes riding on top of an amorphous base pulse; although  deep modulation, down to
to zero intensity, has also been observed \citep{Cordes1990}. 
This so-called microstructure is about 3 orders of magnitude narrower ($\sim\mu$s) than the enveloping subpulses ($\sim$ms), and the spacing between spikes is quasi-periodic.
A linear relation between the periodicity of the microstructure
and the pulsar spin period, 
has been established: first for pulsar spanning the period range $0.15-3.7$\,s \citep{mitra_polarized_2015},
and next  extended to millisecond pulsars \citep{de_detection_2016}. 
Recent observations of the 5.5-second magnetar XTE J1810$-$197 also generally agree with the linear microstructure-period relationship \citep{maan_expected_2019}.
The relation strongly suggests a physical origin, but the exact mechanisms for
the microstructure remains unclear. There are two main theories: a geometrical one, based on distinct tubes of emitting plasma; and
a temporal one, based on modulation of the radio emission \citep[see the discussion in ][]{mitra_polarized_2015}. 

Even if the FRB sub-component resemblance to pulsar microstructure does not immediately allow for their physical interpretation,
we can still apply the empirical  periodicity-period  relation from \citet{mitra_polarized_2015} to these FRBs.
Their observed sub-component periodicities lead to tentative neutron-star spin periods between 0.3 and 10\,s.

% WIDTH + RELATIONSHIP
In microstructure, an increase of pulsar period is met by a linear increase in the component periodicity, as we have just discussed.
A similar relation with the spin period is seen for the micropulse {\em widths} \citep{Cordes79,kramer_high-resolution_2002}. 
If we, again, consider the different components in \FRB{} as micro-structures, we can also apply this relation.
% In \FRB{}, the average intrinsic component width (i.e., after accounting for the inter-channel dispersion smearing and the finite sampling time), defined as FWTM, is 460$\pm$50\,$\mu$s -- strikingly close to their separation. 
In \FRB, the average intrinsic component width (i.e., after accounting for the inter-channel dispersion smearing and the finite sampling time), defined as the full width at half maximum (FWHM), is\,250$\pm$30\,$\mu$s, suggesting a tentative NS spin period in the range 0.3$-$0.6\,s.
% This relation holds for regular micropulses.
Given their required intrinsic brightness, FRBs might actually be more closely related to the giant micropulses as seen in the Vela pulsar \citep{Johnston_2001}.
On average, those have even narrower widths \citep{kramer_high-resolution_2002}. 
The bright single pulses from magnetar XTE~J1810$-$197, however, could be giant micropulses;
and still these agree with the standard linear relationship between the normal micropulse width and spin period \citep{maan_expected_2019}. 
There thus appears to be some scatter in these relations.
We do conclude that if \FRB{} consists of such giant micropulses, 
the tentative neutron-star spin period may increase, and fall on the higher side of the above estimated period range.

One characteristic feature of microstructure is that linear and circular components of emission closely follow quasi-periodic total intensity modulation. This is also true for FRB\,20210213A from \citet{the_chimefrb_collaboration_sub-second_2021}, the only one with polarized data presented. %\ab{What about APERTIF FRBs?} \jvl{Don't have CAL ..}
Unfortunately, as mentioned in \S \ref{sec:polcal}, we were not able to obtain a robust Stokes parameters calibration, which would have been useful to compare to typical pulsar microstructure features.

\subsubsection{The magnetar connection}
The microstructure that was previously known only in pulsars,
has recently also been observed in radio emission from magnetars, especially in the radio-loud magnetar XTE\,J1810-197.
This suggests a further expansion into FRBs is worth considering.
The 2018 outburst of XTE\,J1810-197 led to high-resolution detections and studies, at a number of epochs and radio frequencies \citep{levin_spin_2019, maan_expected_2019, caleb_radio_2021}.
The magnetar exhibits very bright, narrow pulses; and for discussions on the classification of these, we refer to \citet{maan_expected_2019} and \citet{caleb_radio_2021}. 
The source also exhibits multi-component single pulses that show quasi-periodicity is the separation between different components.
If we assume FRBs originate from magnetars, it is possible that the quasi-periodicity that is seen in the sample presented in this paper can be tied to a similar underlying emission mechanism, on a different energy scale. 
Much of the available energy budget is determined by the period and magnetic field. 
The spin-period range we find covers the periods of the known magnetars.
Magnetars with a relatively short period, of about $\gtrsim2$\,s, plus a strong surface magnetic field of $>10^{14}$\,G,
would produce a larger vacuum electric field (due to the rotating magnetic field) than normal, slower, magnetars. 
This field in turn generates stronger particle acceleration and larger Lorentz factors, as required for  FRB luminosities. 
It would be interesting to compare the FRB luminosity of sources with different supposed spin periods, in this model.
%, as well as FRBs  to 1E 1547.0-5408, a 2-s magnetar with the surface magnetic field of $2.22\times10^{14}$\,G, which had recently FRB-like radio bursts \citep{Israel2021}, however such analysis is outside the scope of this work. 

% \ab{I would make the last sentence a separate paragraph and expand more on it, as it is a completely separate theory}

%P0 for R3 lies between 0.8 and 2\,s. It is yet unknown why no such periodicity has been found during direct searches \citep{}. Perhaps the viewing geometry is such that the onpulse window is large and the pulse jitter is comparable to spin period.
%\citet{Li2021} recently made a model of R3's observed 16.3-day period can then be precession period with an offset dipole emission comes outside conventional polar cap). 

%It is possible that FRBs come from the closed magnetospheric field lines, thus resulting in large duty cycle and hindering spin period searches. AB: closed field line theory implies starquakes, not regular polar gap setup.

\subsection{A new FRB morphological type}

Based on the first CHIME/FRB catalog, \cite{pleunis_fast_2021} identified four FRB types based on their morphology. These classes are simple broadband, simple narrowband, temporally complex and downward drifting. Downward drifting bursts are commonly associated with repeating FRBs \citep{hessels_frb_2019}, and A17 and A53 are unequivocal examples of this morphological type. Bursts classified as temporally complex are those presenting more than one component peaking at similar frequencies, but with no constraint on the separation between components. This class makes up for ~5\% of the FRBs presented in the CHIME/FRB catalog, but $\lesssim0.5\%$ of the FRBs show five or more components. 

With the detection of \FRBperiodic, \FRBfivecomp{} and \FRBsixcomp, \cite{the_chimefrb_collaboration_sub-second_2021} propose the existence of a new group of FRBs with periodic pulse profiles. \FRB{} is the first source detected by an instrument other than CHIME/FRB showing a quasi-periodic pulse profile, thus adding a new member to this nascent FRB morphological type. It is not clear yet if these sources are produced by the same progenitors as the other types of FRBs or not. 

All three \FRB, and the \chimefrbs{} have $\leq6$ components, and none of these FRBs reach the $3\sigma$ significance threshold for the periodicity. 
% \sms{added backslash to force space after chime frbs}\sms{It now reads as if none of the components reach 3 sigma in stead of none of the pulses reach 3 sigma in their sub pulse periodicity}
This could be explained by the structure of the magnetosphere of magnetars, formed by emission regions roughly equally spaced. The higher number of components in \FRBperiodic{} might reduce the effect of jitter in each single components and thus increase the significance of the periodicity. Alternatively, given its longer period, envelope duration, and higher periodicity significance, \FRBperiodic{} could have been produced by a magnetar outburst lasting several rotations or a single burst comprising crustal oscillations.  %\jvl{I would be careful with calling this a class. In pulsars we see a \underline{continuum} of behaviour that defies classes. Not every outlier is a class.. they are all one big class, I'd say.}

Considering the scarcity of FRBs with a regular separation between components, the detection of \FRB{} is remarkable, given the number of events detected with Apertif is a few orders of magnitude lower than CHIME/FRB.
This suggests that a larger fraction of FRBs with quasi-periodic components might be visible at higher frequencies. 
Certain astrophysical and instrumental effects could play a role at blurring together multi-component bursts at lower frequencies. This includes interstellar scattering, which evolves as $\tau\sim\nu^{-4}$, as well as intra-channel dispersive smearing, which scales as $\nu^{-3}$ and finite sampling of existing instruments that prevent detection of components narrower than the sampling interval. Searches of multi-component FRBs with a regular spacing might thus be more successful at higher frequencies.

The bursts A17 and A53 from \RIII{} discussed in this work, although similar by eye to \FRB, do not show conclusive evidence for periodicity in their subcomponents. This, added to the downward drift in frequency of the subcomponents, common in bursts from repeating FRBs, differs from \FRB{} and the CHIME/FRB \FRBperiodic, \FRBfivecomp, and \FRBsixcomp. This might suggest the presence of a different emission process between the bursts from repeating FRBs and the FRBs with quasi-periodic components or an additional mechanism modulating the bursts.

%--------------------------------------

\section{Conclusions}
\label{sec:conclusions}

In this paper, we have presented a new fast radio burst detected with Apertif, \FRB. This FRB shows a quasi-periodic
structure analogous to the three FRBs presented in \cite{the_chimefrb_collaboration_sub-second_2021}.
The average spacing between its five components, however, is $\sim0.415$\,ms, in the sub-millisecond regime. We have
performed a timing analysis of the FRB, and we conclude that the significance of the periodicity is $\sim2.5\sigma$,
below the standard $3\sigma$ threshold, but comparable to the significance of the \chimefrbs. Given the scarcity of FRBs with quasi-periodic structure detected so far, they have been likely produced via the same mechanism, and we postulate that they constitute a new FRB morphological type.

Additionally, we have performed a timing analysis of bursts A17 and A53 from \cite{pastor-marazuela_chromatic_2021}, the bursts with the highest number of visible subcomponents. 
We find no conclusive evidence of periodicity in these bursts, although the periodicity significance of A17 is above $3\sigma$ according to the \sshat{} test. In A53, a burst with $\geq11$ components, we find an average subcomponent spacing of $\sim1.7$\,ms. For the five-component burst A17, the average subcomponent separation is $\sim1$\,ms. These timescales are of the same order of magnitude as the subcomponent separation in \FRB{}. 

We have discussed several interpretations of the quasi-periodicity, and we rule out the sub-millisecond pulsar and the binary compact object merger scenarios as potential progenitors to \FRB. However, its morphology is comparable to pulsar subpulses and pulsar microstructure, and similar structures have been observed in single pulses from radio-loud magnetars. We thus conclude that the structure in the magnetosphere of a magnetar could be at play in producing bursts such as \FRB{} and the \chimefrbs. On the other hand, given the high significance of the period detected in \FRBperiodic, as well as the higher burst separation of $\sim200$\,ms, the structure seen in this CHIME/FRB burst could have been produced by a magnetar outburst lasting several NS rotations. 

\begin{acknowledgements}
We thank Kendrick Smith for helpful conversations. 

This research was supported by 
the European Research Council under the European Union's Seventh Framework Programme
(FP/2007-2013)/ERC Grant Agreement No. 617199 (`ALERT'),
by Vici research programme `ARGO' with project number
639.043.815, financed by the Dutch Research Council (NWO), and 
through CORTEX (NWA.1160.18.316), under the research programme NWA-ORC, financed by NWO.
Instrumentation development was supported 
by NWO (grant 614.061.613 `ARTS') and the  
Netherlands Research School for Astronomy (`NOVA4-ARTS', `NOVA-NW3', and `NOVA5-NW3-10.3.5.14').
PI of aforementioned grants is JvL.
EP further acknowledges funding from an NWO Veni Fellowship.
SMS acknowledges support from the National Aeronautics and Space Administration (NASA) 
  under grant number NNX17AL74G issued through the NNH16ZDA001N Astrophysics Data Analysis Program (ADAP).
DV and AS acknowledge support from the Netherlands eScience Center (NLeSC) under grant ASDI.15.406.
KMH acknowledges financial support from the State Agency for Research of the Spanish MCIU through the "Center of Excellence Severo Ochoa" award to the Instituto de Astrof\'{i}sica de Andaluc\'{i}a (SEV-2017-0709) and from the coordination of the participation in SKA-SPAIN, funded by the Ministry of Science and innovation (MICIN).
JMvdH and KMH acknowledge funding from the European Research Council under the European Union’s Seventh Framework Programme (FP/2007-2013)/ERC Grant Agreement No. 291531 (‘HIStoryNU’).

This work makes use of data from the Apertif system installed at the Westerbork Synthesis Radio Telescope owned by ASTRON. ASTRON, the Netherlands Institute for Radio Astronomy, is an institute of NWO.
This research has made use of the NASA/IPAC Extragalactic Database, which is funded by the National Aeronautics and Space Administration and operated by the California Institute of Technology.
\end{acknowledgements}

% WARNING
%-------------------------------------------------------------------
% Please note that we have included the references to the file aa.dem in
% order to compile it, but we ask you to:
%
% - use BibTeX with the regular commands:
%\bibliographystyle{aa} % style aa.bst
\bibliographystyle{yahapj}
\bibliography{biblio}

% - join the .bib files when you upload your source files
%-------------------------------------------------------------------

\begin{appendix} %First appendix
\section{Galaxies within FRB error region} \label{app:galaxies}

In this appendix we give Table~\ref{tab:galaxies} with the galaxies identified within the 99\% confidence interval on the localisation of \FRB.

\begin{table}[h]
    \centering
    \caption{Galaxies within the error region of \FRB.}
    \begin{tabular}{c c}
    \hline\hline
    Object name & Magnitude and filter \\
    \hline
    SDSS J135123.63+490313.1 & 21.9g \\
    %WISEA J135123.74+490236.6
    SDSS J135123.70+490236.7 & 22.3g \\
    SDSS J135124.48+490159.8 & 22.8g \\
    %WISEA J135124.59+490131.9 
    SDSS J135124.60+490131.7 & 23.5g \\
    SDSS J135125.72+490128.7 & 22.1g \\
    \hline
    \end{tabular}
    \tablefoot{These galaxies were queried from the NED database. None have a measured redshift.}
    \label{tab:galaxies}
\end{table}

\section{2D auto-correlation functions of \FRB, A17 and A53} \label{app:2D_ACF}

In this section we show the two-dimensional auto-correlation functions (AFC)\footnote{Computed with the \texttt{signal.correlate2d} function of the SciPy python package.} of \FRB, A17 and A53. After computing the ACFs they were fitted to two-dimensional gaussians with an inclination angle. This inclination angle gives us a robust drift-rate estimate of the burst subcomponents \cite{hessels_frb_2019}. We obtained the time and frequency ACFs by averaging in frequency and time respectively. 

The FWHM of a lorentzian fitted to the central peak of the frequency ACF gives the scintillation bandwidth of the FRB.
In addition, any structure in the temporal ACF can provide with information about any dominant timescales in the pulse structure.

\begin{figure}[h]
    \centering
    \includegraphics[width=\hsize]{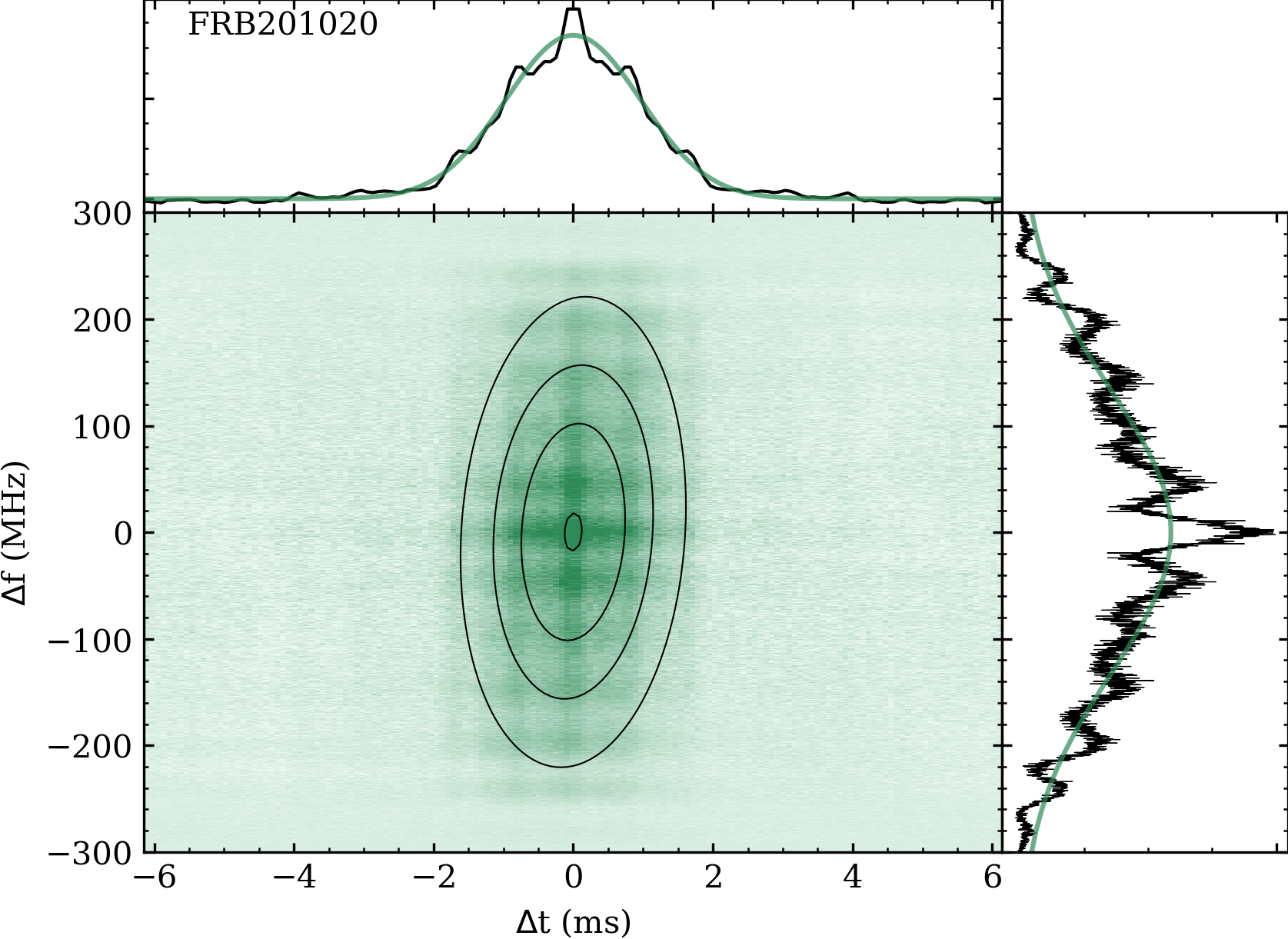}\\
    \includegraphics[width=\hsize]{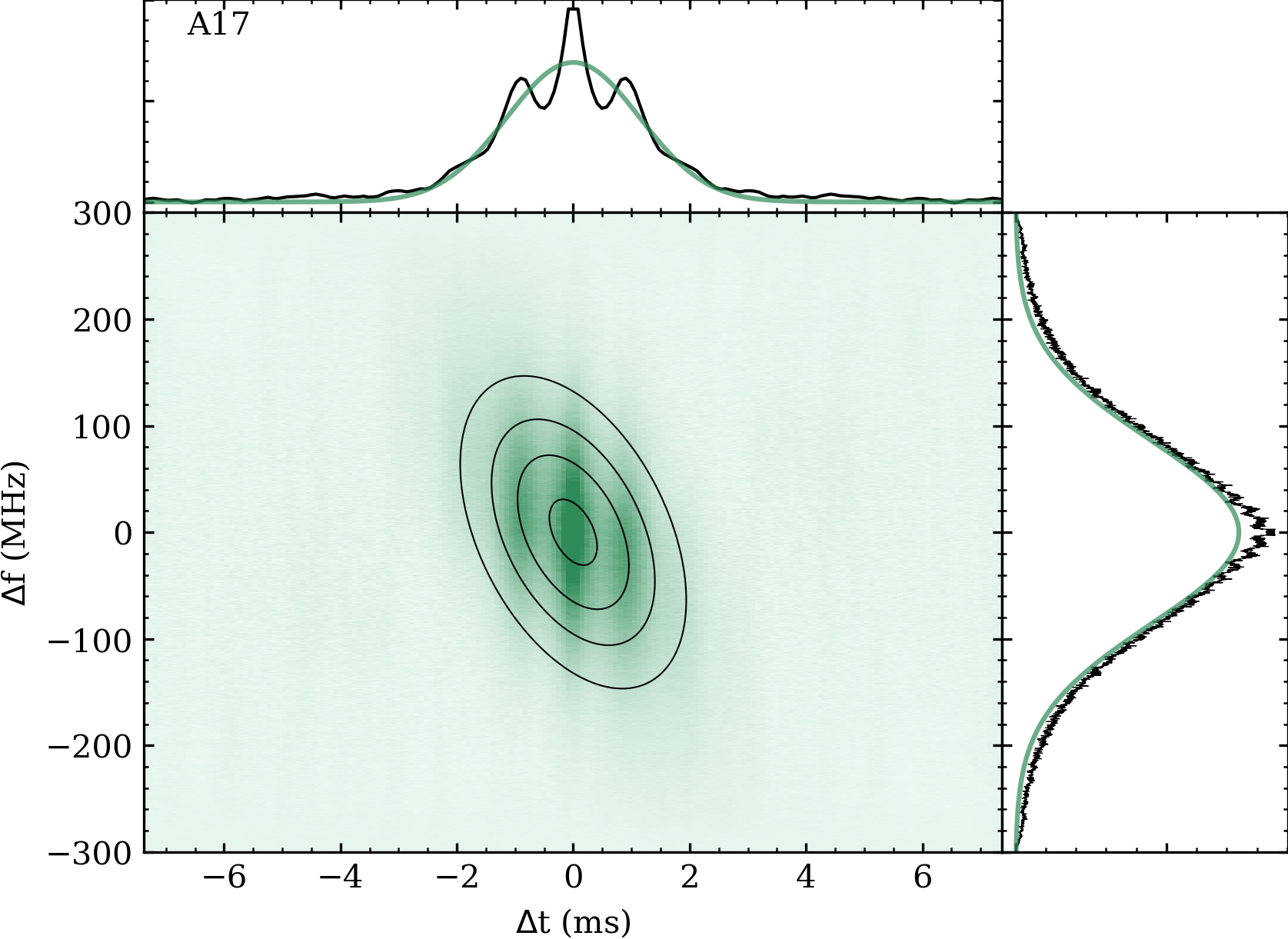}\\
    \includegraphics[width=\hsize]{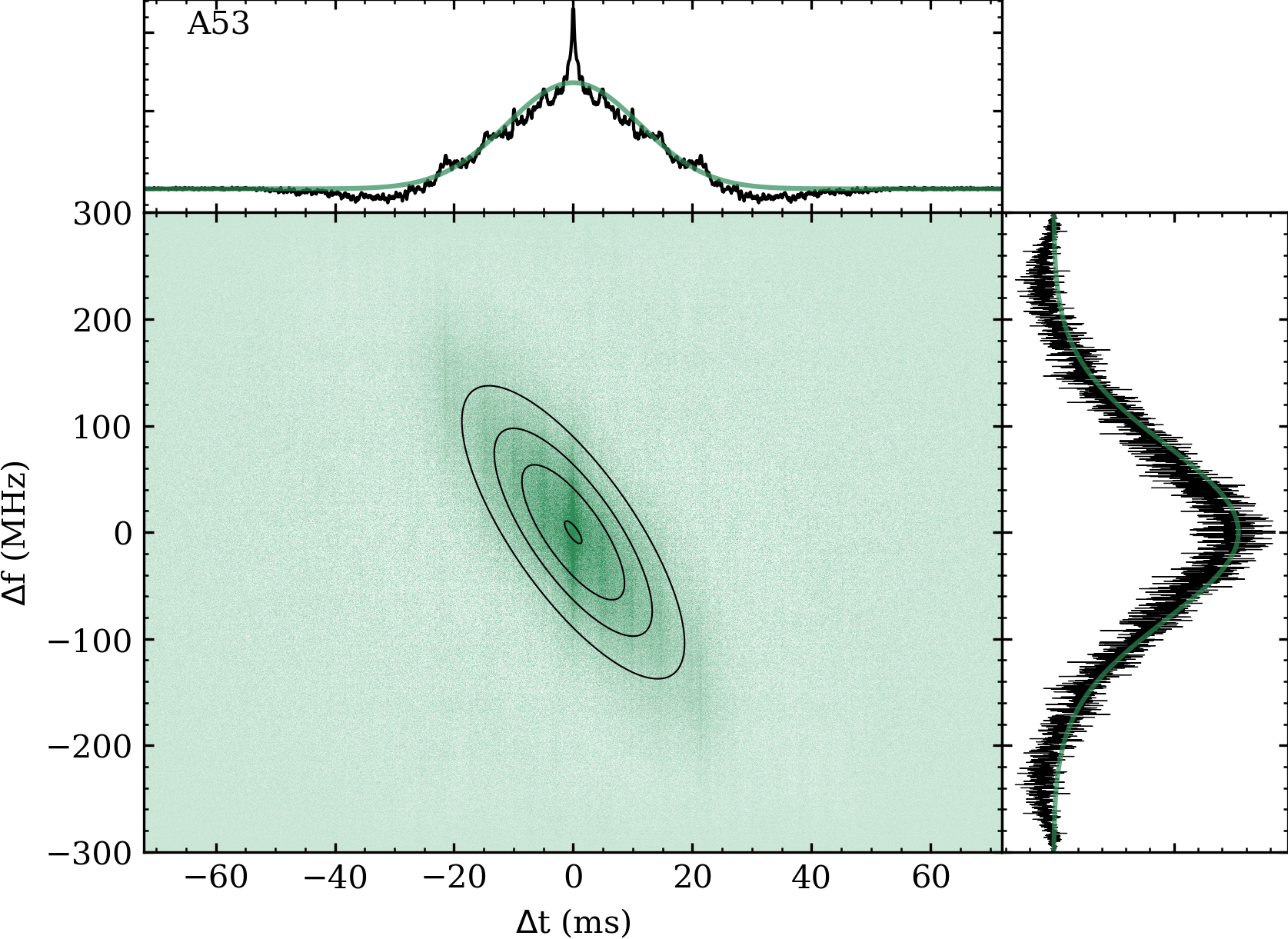}
    \caption{2D ACFs of \FRB, A17 and A53 from top to bottom. The lower left panel of each bursts shows the 2D ACF in green, and the 2D gaussian fit as black elliptical contours. The top panel of each bursts shows the frequency-averaged temporal ACF in black and the gaussian fit in green. The left panels show the time-averaged frequecial ACF in black and the gaussian fit in green.}
    \label{fig:acf2d}
\end{figure}

\section{Rotation measure}
\label{app:rm}

\begin{figure}[h]
    \centering
    \includegraphics[width=\hsize]{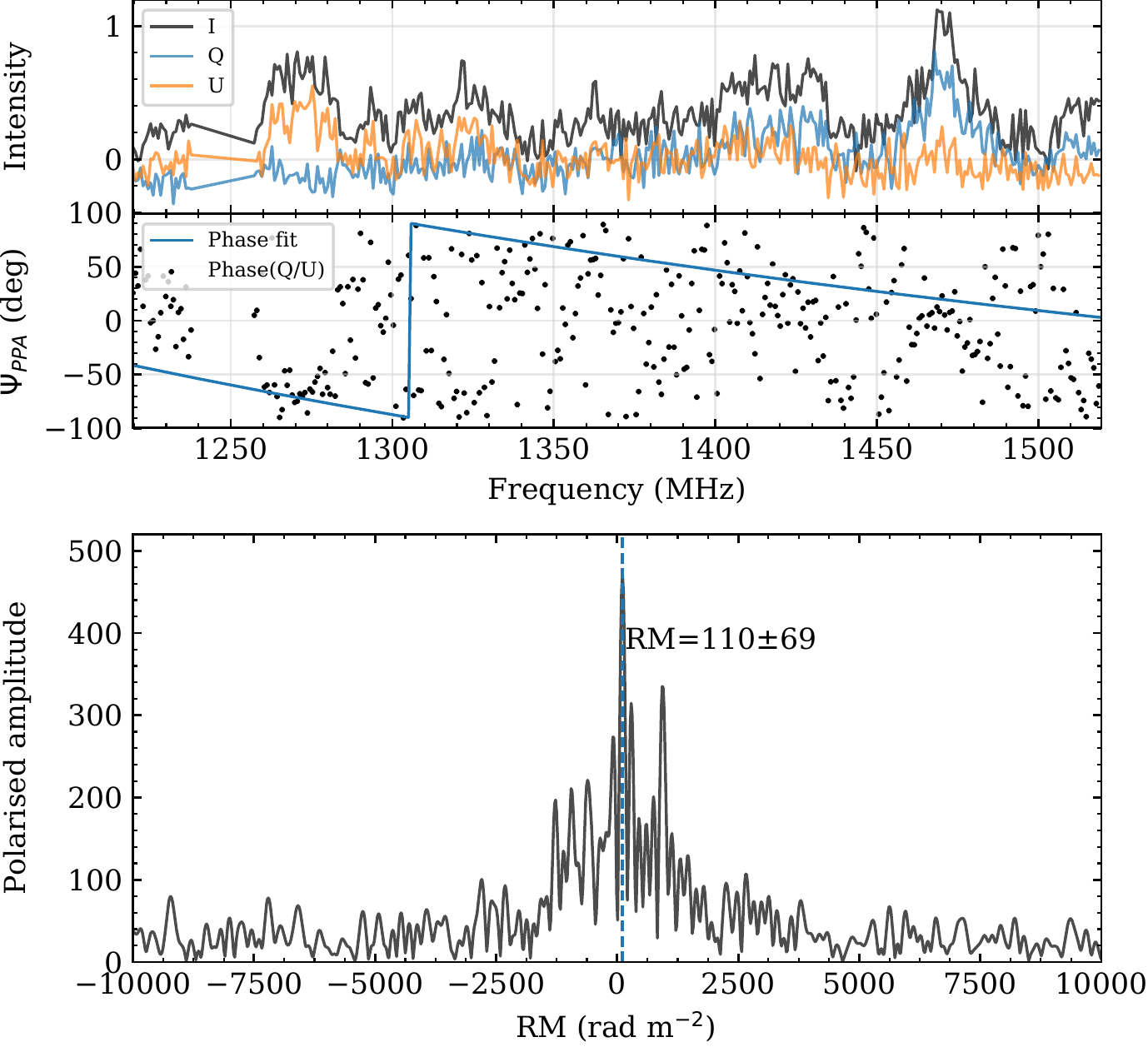}
    \caption{Measured polarisation properties of \FRB. The top panel shows the Stokes parameters I (black), Q (blue) and U (orange) as a function of frequency, calibrated by assuming Stokes V to be zero. The second panel shows the measured phase between Q and U (black dots) and its fit. The bottom panel shows the RM-synthesis solution, with the polarised amplitude as a function of RM. The dashed blue line marks the position of the maximum, at RM=$+110\pm69$\pccm.}
    \label{fig:rm}
\end{figure}

In this section we show the results from the polarisation and Faraday rotation calibration, shown in Fig.~\ref{fig:rm}. Note that the best RM of $110\pm69$\pccm is close to the lower limit of what Apertif can measure given its central frequency and bandwidth.

\section{Maximal number of orbits in binary merger} \label{app:dphi_max}

To get the expression for the maximal orbital phase a binary system can complete after reaching a certain frequency, $\Delta\phi_{\text{max}}$ as given in Equation~\ref{eq:dphi_max}, we start with Eq.~16 from \cite{cutler_gravitational_1994}:

\begin{equation}
    \frac{df}{dt} = \frac{96}{5}\pi^{8/3} \left( \dfrac{G\mathcal{M}}{c^3} \right)^{5/3} f^{11/3} = \mathbb{M} f^{11/3}.
\end{equation}

We define $\mathbb{M}\equiv \frac{96}{5}\pi^{8/3}\left( \frac{G\mathcal{M}}{c^3}\right)^{5/3}$, where $\mathcal{M}\equiv\mu^{3/5}M^{2/5}$ is the chirp mass, $M=M_1+M_2$ the total mass of the system and $\mu=M_1M_2/M$ the reduced mass. $G$ is the gravitational constant and $c$ the speed of light in the vacuum (note that we do not make the replacement $G=c=1$). We solve the differential equation to find $f(t)$:

\begin{equation}
    f(t) = \left( f_0^{-8/3} - \frac{8\mathbb{M}}{3} t \right)^{-3/8}.
\end{equation}

Here, $f_0$ is the frequency at $t=0$. To get the orbital phase, we solve the differential equation $\frac{d\phi}{dt}=2\pi f(t)$ and we get:

\begin{equation}
    \phi(t) = -\dfrac{6\pi}{5\mathbb{M}} \left(f_0^{-8/3} - \frac{8\mathbb{M}}{3}t\right)^{5/8} + \mathcal{C}.
\end{equation}

The value of $\mathcal{C}$ is given by the initial conditions; $\phi(0)=0$ so we get $\mathcal{C}=\frac{6\pi}{5\mathbb{M}} f_{0}^{-5/3}$. In the simplest scenario where we assume the objects in the binary system to be point-like sources, the merger will happen when $\frac{8\mathbb{M}}{3}t - f_0^{-8/3} = 0$, from which we define the maximal orbital phase $\Delta\phi_{\text{max}} = \mathcal{C}$. Finally, we get:

\begin{equation}
    \Delta\phi_{\text{max}} = \dfrac{1}{16\pi^{5/3}} \left( \dfrac{G\mathcal{M}}{c^3} \right)^{-5/3} f_0^{-5/3}.
\end{equation}

Since we take our initial frequency to be the frequency corresponding to the FRB subcomponent separation, we get Eq.~\ref{eq:dphi_max}.

\end{appendix}

\end{document}